%% file: paper1.tex
\documentclass[12pt]{article}
\usepackage{a41}
\usepackage{a41}
\usepackage{cite}
\usepackage{amsmath}
\usepackage{amssymb}
\usepackage{color}  

\usepackage[rflt]{floatflt}              
\usepackage{float}
\usepackage{slashed}
\def\b0{\beta_0}

\usepackage{array}

\newcommand{\HA}{{\rm H}}
\newcommand{\HAA}{\tilde{\rm H}}
\newcommand{\sigm}{\tilde{\sigma}}

\usepackage[english]{babel}
\usepackage[latin1]{inputenc}
\usepackage[T1]{fontenc}
\usepackage{ae}

\usepackage{url}

\usepackage{amsmath, amsthm, amssymb}
\newtheorem{thm}{Theorem}[section]

\newtheorem{definition}[thm]{Definition}

 \newcommand{\GeV}{\mathrm{GeV}}
 \newcommand{\keV}{\mathrm{keV}}
 \newcommand{\MeV}{\mathrm{MeV}}

 \newcommand{\MS}{\overline{\sf MS}}

\newcommand{\Li}{{\rm Li}}

\newcommand{\Mvec}{{\rm\bf M}}

\newcommand{\ep}{\varepsilon}

\usepackage{rotating}

\usepackage{graphicx}
\newcounter{mmacnt}
\def\restartmma{\setcounter{mmacnt}{0}}
\restartmma \catcode`|=\active
\def|#1|{\mathrm{#1}}
\catcode`|=12
\newenvironment{mma}{
 \par\smallskip
 \catcode`|=\active
 \parskip=0pt\parindent=0pt % locally
 \small
 \def\In##1\\{%
\def\linebreak{\hfill\break\null\qquad}%
\refstepcounter{mmacnt}
\hangindent=2.5em\hangafter=0
\leavevmode
\llap{\tiny\sffamily n[\arabic{mmacnt}]:=\kern.5em}%
\mathversion{bold}\footnotesize$\displaystyle##1$\normalsize
\mathversion{normal}\par
 }%
 \def\Print##1\\{%
\def\linebreak{\hfill\break}%
\hangindent=2.5em\hangafter=0
\leavevmode ##1\par}%
 \def\Out##1\\{%
\def\linebreak{$\hfill\break\null\hfill$}%
\kern\abovedisplayskip\par
\hangindent=2.5em\hangafter=0
\leavevmode
\llap{\tiny\sffamily Out[\arabic{mmacnt}]=\kern.5em}
\footnotesize$\displaystyle##1$\normalsize\hfill\null\par
\kern\belowdisplayskip
 }%
 \def\Warning##1##2\\{%
\def\linebreak{\hfill\break}%
\hangindent=2.5em\hangafter=0
\leavevmode
{\scriptsize##1 : ##2}\par}%
}{%
 \par\smallskip
}

\usepackage{color}

\newenvironment{fshaded}{%
\MakeFramed {\FrameRestore}
}%
{\endMakeFramed}

\def\b0{\beta_0}

\def\Gp0{{\Gamma^{'}_0}}
\def\Gp1{{\Gamma^{'}_1}}
\def\Gp2{{\Gamma^{'}_2}}

\allowdisplaybreaks[4]

\begin{document}
\setlength{\baselineskip}{0.515cm}

\sloppy
\thispagestyle{empty}
\begin{flushleft}
DESY 19--231
%  \hfill {\tt arXiv:2003.xxxxx[hep-ph]}
\\
DO-TH 19/32\\
TTP20--012\\
SAGEX 19--36\\
\end{flushleft}

\mbox{}
\vspace*{\fill}
\begin{center}

{\LARGE\bf Subleading Logarithmic QED \boldmath Initial} 

\vspace*{2mm}
{\LARGE\bf \boldmath State Corrections   to $e^+e^- \rightarrow \gamma^*/{Z^{0}}^*$} 

\vspace*{2mm}
{\LARGE\bf \boldmath to $O(\alpha^6 L^5)$}

\vspace{3cm}
\large
{\large 
J.~Ablinger$^a$, 
J.~Bl\"umlein$^b$, 
A.~De~Freitas$^b$,
and K.~Sch\"onwald$^{b,c}$
}

\vspace{1.cm}
\normalsize
{\it $^a$~Research Institute for Symbolic Computation (RISC),\\
Johannes Kepler University, Altenbergerstra\ss{}e 69, A--4040, Linz, Austria}

\vspace*{2mm}
{\it   $^b$~Deutsches Elektronen--Synchrotron, DESY,}\\
{\it   Platanenallee 6, D--15738 Zeuthen, Germany}

\vspace*{2mm}
{\it  $^c$~Institut f\"ur Theoretische Teilchenphysik,\\
Karlsruher Institut f\"ur Technologie (KIT) D-76128 Karlsruhe, Germany}

%%\today

\end{center}
\normalsize
\vspace{\fill}
\begin{abstract}
\noindent
Using the method of massive operator matrix elements, we calculate the subleading QED initial state radiative 
corrections to the process $e^+e^- \rightarrow \gamma^*/Z^*$ for the first three logarithmic contributions from 
$O(\alpha^3 L^3), O(\alpha^3 L^2), O(\alpha^3 L)$ to $O(\alpha^5 L^5), O(\alpha^5 L^4), O(\alpha^5 L^3)$ and 
compare their effects to the leading contribution $O(\alpha^6 L^6)$ and one more subleading term $O(\alpha^6 L^5)$. 
The calculation is performed in the limit of large center of mass energies squared $s \gg m_e^2$. These terms 
supplement the known corrections to $O(\alpha^2)$, which were completed recently. Given the high precision at 
future colliders operating at very large luminosity, these corrections are important for concise theoretical 
predictions. The present calculation needs the calculation of one more two--loop massive operator matrix element 
in QED. The radiators are obtained as solutions of the associated Callen--Symanzik equations in the massive case. 
The radiators can be expressed in terms of harmonic polylogarithms to weight {\sf w = 6} of argument $z$ and $(1-z)$ 
and in Mellin $N$ space by generalized harmonic sums. Numerical results are presented on the position of the $Z$ 
peak and corrections to the $Z$ width, $\Gamma_Z$. The corrections calculated result into a final theoretical 
accuracy for $\delta M_Z$ and $\delta \Gamma_Z$ which is estimated to be of $O(30~\keV)$ at an anticipated 
systematic accuracy at the FCC\_ee of $\sim 100~\keV$. This precision cannot be reached, however, by including only 
the corrections up to $O(\alpha^3)$.
\end{abstract}

\vspace*{\fill}
\noindent
% \numberwithin{equation}{section}
%%%%%%%%%%%%%%%%%%%%%%%%%%%%%%%%%%%%%%%%%%%%%%%%%%%%%%%%%%%%%%%%%%%%%%%%%%%%%%%%%%%%%%%%%%%%%%%%%%%%%%%%%%%%%%%%%%%%%%%%%%%%%%%%%%%
\newpage 

%--------------------------------------------------------------------------------------------------------
\section{Introduction}
\label{sec:1}
%--------------------------------------------------------------------------------------------------------

\vspace*{1mm}
\noindent
At the planned future $e^+e^-$ facilities which operate at high energy and at large luminosity, like  
the ILC, CLIC \cite{ILC,Aihara:2019gcq,Mnich:2019,CLIC}, the FCC\_ee \cite{FCCEE}, and also muon colliders
\cite{Delahaye:2019omf}, tests of the Standard Model are possible at unprecedented accuracy. This concerns
a further detailed exploration of the $Z$ peak, beyond what was possible at LEP \cite{ALEPH:2005ab}, 
detailed studies of Higgs boson production using $Z H$ final states, accurate scans of the top-quark
threshold region, and various other precision measurements. Due to this the accuracy of the masses and 
widths of the heaviest particles of the Standard Model will be significantly improved.

One important ingredient to these experimental precision studies are the QED radiative corrections and in 
particular those due to initial state radiation (ISR). Very recently the $O(\alpha^2)$ ISR corrections for the 
process $e^+e^- \rightarrow \gamma^*/Z^*$ have been completed in a direct calculation~\cite{Blumlein:2019srk,
Blumlein:2019pqb,QED2019}. Here $\alpha$ denotes the fine structure constant. It has been shown that the 
result for all channels of a previous calculation 
\cite{Berends:1987ab} needed to be corrected in the non--logarithmic terms at $O(\alpha^2)$. Agreement has 
been found with the results of Ref.~\cite{Blumlein:2011mi}. Due to this it has also been proven that one may 
use the method of massive operator matrix elements (OMEs) for these calculations and that the Drell--Yan process
factorizes for massive fermionic states.

In the present paper we take advantage of this method and extend the calculation to the first three 
logarithmic terms up to the order $O(\alpha^5)$  beyond the complete $O(\alpha^2)$ corrections and thus reach
$O(\alpha^5 L^3)$. For comparison we also calculate the leading order contributions of $O(\alpha^6 L^6)$ and one
more subleading term $O(\alpha^6 L^5)$, with
$L = \ln(s/m_e^2)$, $s$ denotes the center of mass energy 
squared of the annihilation process and $m_e$ is the electron mass. The universal corrections $O((\alpha L)^k)$ 
are known in analytic form to order  $k = 5$, cf.~\cite{Skrzypek:1992vk,Jezabek:1992bx,Przybycien:1992qe,
Blumlein:1996yz,Arbuzov:1999cq,Arbuzov:1999uq,Blumlein:2004bs,Blumlein:2007kx}, accounting for the non--singlet 
and singlet contributions in the unpolarized and polarized case. The method used in 
Ref.~\cite{Blumlein:2011mi} can be extended to higher orders in the coupling constant. For the first 
subleading term at $O(\alpha^3)$\footnote{Subleading corrections to differential cross sections have also been 
studied, cf.~\cite{Blumlein:2002fy}.}, all ingredients forming the 
radiators in terms of Mellin transforms are known from the calculation of the anomalous dimensions in QCD 
\cite{NLOand}, the Wilson coefficients of the massless Drell--Yan process 
\cite{Hamberg:1990np,Harlander:2002wh,Blumlein:2019srk} and the massive OMEs in \cite{Blumlein:2011mi}. To 
obtain the $O(\alpha^3 L)$ and $O(\alpha^3 L)$ correction we need also to calculate the massive OME 
$\Gamma^{(1)}_{\gamma e}$.
This also applies to the higher order subleading contributions. This series could be continued straightforwardly to 
higher and higher order, for the first three terms at each order in the logarithmic expansion in Mellin $N$ space 
on the expense of longer and longer expressions.

It turns out that it is indeed the case that one has to reach at least corrections of the order $O(\alpha^5 L^4)$ 
for the ISR 
corrections to satisfy the ambitions goals at the FCC\_ee of $\sim 100$~keV both for the $Z$ mass, $M_Z$, and the $Z$ 
width, $\Gamma_Z$, on the theoretical side. The method of massive operator matrix elements \cite{Blumlein:2011mi} makes 
this calculation possible. In the present approach the constant term of $O(\alpha^3)$ and its higher order logarithmic
extensions are still missing. They require still higher order massive OMEs and also massless Wilson 
coefficients in analytic form. Their size is, however, gradually smaller. 

The paper is organized as follows. In Section~\ref{sec:2} we present the structure of the QED ISR radiative 
corrections to the Born cross section following from the renormalization group equations (RGEs), for which we present 
the analytic solutions in terms of anomalous dimensions, massive OMEs and massless Wilson coefficients for all 
contributions calculated in the present paper. In Section~\ref{sec:3} we calculate the massive OME 
$\Gamma^{(1)}_{\gamma e}$. The radiator functions in $z$ space for the $O(\alpha^3)$ and $O(\alpha^4)$ 
corrections are presented 
in Section~\ref{sec:4}.  The radiators can be expressed in terms of harmonic polylogarithms \cite{Remiddi:1999ew}, 
to which the Nielsen integrals and classical polylogarithms form a subset \cite{NIELSEN,CLPOLY}. 
In Section~\ref{sec:5}, we present numerical results on the corrections in the kinematic
region around the $Z$ peak and we determine the corresponding corrections to the pole mass of the 
$Z$ boson and the $Z$ 
boson width, $\Gamma_Z$. Section~\ref{sec:6} contains the conclusions. In Mellin $N$ space the radiators are given in 
terms of harmonic and generalized harmonic sums. If compared to the $z$ space representation they are more compact.
Due to the appearance of generalized harmonic sums we derive in Appendix~\ref{sec:A} the singularity structure of 
the radiators in the complex $N$-plane and in Appendix~\ref{sec:B} we present all radiators 
calculated in the present paper in Mellin space.
%--------------------------------------------------------------------------------------------------------
\section{The Initial State Corrections to the \boldmath $e^+e^-$ Annihilation Cross Section}
\label{sec:2}
%--------------------------------------------------------------------------------------------------------

\vspace*{1mm}
\noindent
The initial state QED radiative corrections to $e^+e^- \rightarrow \gamma^*/Z^*$ can be calculated 
by applying the method of massive operator matrix elements, cf.~\cite{Blumlein:2011mi}. It  has been demonstrated
by the recent complete calculation in \cite{Blumlein:2019srk,Blumlein:2019pqb,QED2019} that the effective method 
of Ref.~\cite{Blumlein:2011mi} is delivering the complete result. Particular non--logarithmic contributions with 
vanishing massive OME to the Drell--Yan process in the constant term $O(\alpha^2)$ could be structurally absorbed 
in the relations and appear as contributions to the massless Wilson coefficients given in 
\cite{Blumlein:2011mi},~Eqs.~(40--42). It is due to this agreement, that one 
can now safely apply this method also for subleading logarithms to higher orders in the coupling constant by 
solving the associated renormalization group equations in the massive case.
The method has been used in Ref.~\cite{Berends:1987ab} before for the logarithmic enhanced contributions
to $O(\alpha^2 L)$.\footnote{Note 
the necessary corrections of relations in \cite{Berends:1987ab} given in Ref.~\cite{Blumlein:2011mi}.}

The massive effects due to the finite electron mass $m_e$ enter here through process--independent massive operator 
matrix 
elements. The Wilson coefficients are those of the massless Drell--Yan process \cite{Hamberg:1990np,Harlander:2002wh,
Blumlein:2019srk}. Note that there are differences between the corrections to the vector and axial--vector coupling. 

In $z$--space the different contributions to the radiators are connected by Mellin convolutions $\otimes$ which 
are defined by
%--------------------------------------------------------------------------------------------------------
\begin{eqnarray}
[A \otimes B](z) = \int_0^1 dx_1 \int_0^1 dx_2 \delta(1 - x_1 x_2) A(x_1) B(x_2).
\end{eqnarray}
%--------------------------------------------------------------------------------------------------------
The Mellin transform reads
%--------------------------------------------------------------------------------------------------------
\begin{eqnarray}
\label{eq:MELTRA}
\Mvec[f(z)](N) = \int_0^1 dz z^{N-1} f(z),~~~~
\Mvec[[f(z)]_+](N) = \int_0^1 dz (z^{N-1}-1) f(z)
\end{eqnarray}
%--------------------------------------------------------------------------------------------------------
for regular functions and +-distributions, respectively.
The most recently calculated quantities are the massive OMEs given in \cite{Blumlein:2011mi}.

The radiative corrections to the differential scattering cross section is given by
%--------------------------------------------------------------------------------------------------------
\begin{eqnarray}
\label{EQ:SI1}
\frac{d\sigma_{e^+e^-}}{ds'} = \frac{1}{s} \sigma^{(0)}\left\{1 + \sum_{k = 1}^\infty a_0^k \sum_{l=0}^k c_{k, l} 
{\rm \bf L}^{k-l} \right\}
\end{eqnarray}
%--------------------------------------------------------------------------------------------------------
in $z$ space. Here we defined
%--------------------------------------------------------------------------------------------------------
\begin{eqnarray}
{\rm \bf L} = 
L  + \ln(z),~~~~~L = \ln \left(\frac{s}{m_e^2}\right),
\end{eqnarray}
%--------------------------------------------------------------------------------------------------------
and $\sigma^{(0)}$ denotes the Born cross section, cf.~\cite{Blumlein:2011mi},~Eq.~(8), and Ref.~\cite{BDJ}.
$a_0 = a(\mu^2 = m_e^2)$ is the normalized fine structure constant with $a(\mu^2) = \alpha(\mu^2)/(4\pi)$, which 
we will widely use in the following. Here it is convenient to refer to $a_0$ only and account also for all 
evolution contributions by the functions $c_{k,l}$.

For the Born cross section for $e^+e^-$ annihilation, $\sigma_0$, we  will consider
$s$-channel $e^+e^-$ annihilation into a virtual gauge boson $(\gamma, Z)$ which decays into a fermion pair
$f \overline{f}$ and $e \neq f$. This process both describes lower energy $\gamma$-exchange and the $Z$-resonance.
%-----------------------------------------------------------------------
\begin{eqnarray}
\label{eq:BO1}
\frac{d\sigma^{(0)}(s)}{d \Omega} &=& \frac{\alpha^2}{4 s}
                                        N_{C,f} \sqrt{1 - \frac{4 m_f}{s}}
\nonumber\\
& & \times
\left[\left(1+ \cos^2\theta + \frac{4 m_f^2}{s} \sin^2\theta \right) G_1(s)
- \frac{8 m_f^2}{s} G_2(s)
+ 2 \sqrt{1-\frac{4m_f^2}{s}} \cos\theta G_3(s)\right]
\nonumber\\                  
&& \times {\cal G}(s)~,
\\
%----------------------
\label{eq:BO2}
\sigma^{(0)}(s) &=& \frac{4 \pi \alpha^2}{3 s}
                                        N_{C,f} \sqrt{1 - \frac{4 m_f}{s}}
\left[\left(1 + \frac{2 m_f^2}{s} \right) G_1(s)
- 6 \frac{m_f^2}{s} G_2(s)\right] {\cal G}(s)~,
\end{eqnarray}
%-----------------------------------------------------------------------
see e.g. \cite{BDJ,WB}\footnote{Note a missing term in \cite{Berends:1987ab}, Eq.~(2.5).}.
Here the final state fermions are considered to be no electrons, to obtain an $s$-channel Born cross section. 
In
Eqs.~(\ref{eq:BO1}, \ref{eq:BO2}) the electron mass is neglected kinematically. $N_{C,f}$ is the number of colors 
of the final state fermion, with $N_{C,f} = 1$ for colorless fermions,
and $N_{C,f} = 3$ for quarks. The function ${\cal G}(s) = 1$ in the case of the pure perturbative calculation.
$s$ is the cms energy, $\Omega$ is the spherical angle, $\theta$ the cms scattering
angle, and the effective couplings $G_i(s)|_{i=1...3}$ read
%-----------------------------------------------------------------------
\begin{eqnarray}
G_1(s) &=& Q_e^2 Q_f^2 + 2 Q_e Q_f v_e v_f {\sf Re}[\chi_Z(s)]
          +(v_e^2+a_e^2)(v_f^2+a_f^2)|\chi_Z(s)|^2,\\
G_2(s) &=& (v_e^2+a_e^2) a_f^2 |\chi_Z(s)|^2, \\
G_3(s) &=& 2 Q_e Q_f a_e a_f {\sf Re}[\chi_Z(s)] + 4 v_e v_f a_e a_f |\chi_Z(s)|^2.
\end{eqnarray}
%-----------------------------------------------------------------------
The reduced $Z$--propagator is given by
%-----------------------------------------------------------------------
\begin{eqnarray}
\chi_Z(s) = \frac{s}{s-M_Z^2 + i M_Z \Gamma_Z},
\end{eqnarray}
%-----------------------------------------------------------------------
where $M_Z$ and  $\Gamma_Z$ are the mass and the width of the $Z$ boson and $m_f$ is the mass of the final
state fermion. $Q_{e,f}$ are the electromagnetic charges of the electron $(Q_e = -1)$
and the final state fermion, respectively, and
the electro--weak couplings $v_i$ and $a_i$ read
%-----------------------------------------------------------------------
\begin{eqnarray}
v_e &=& \frac{1}{\sin\theta_w \cos\theta_w}\left[I^3_{w,e} - 2 Q_e
\sin^2\theta_w\right],\\
a_e &=& \frac{1}{\sin\theta_w  \cos\theta_w} I^3_{w,e}, \\
v_f &=& \frac{1}{\sin\theta_w \cos\theta_w}\left[I^3_{w,f} - 2 Q_f
\sin^2\theta_w\right],\\
a_f &=& \frac{1}{\sin\theta_w  \cos\theta_w} I^3_{w,f}~,
\end{eqnarray}
%-----------------------------------------------------------------------
where $\theta_w$ is the weak mixing angle, and $I^3_{w,i} = \pm 1/2$ the third component
of the weak isospin for up and down particles, respectively.

The inclusive $s$--channel annihilation scattering cross section $\sigma_{e^+ e^-}$ is given by
%--------------------------------------------------------------------------------------------------------
\begin{eqnarray}
\label{eq:incl}
\sigma_{e^+ e^-}(s) &=& \int_{s_0}^s ds' \frac{d \sigma_{e^+ e^-}(s')}{ds'}
= \int_{z_0}^1 dz \sigma^{(0)}(s z) R(z,s/m_e^2),
\end{eqnarray}
%--------------------------------------------------------------------------------------------------------
with $s' = sz$, $\sigma^{(0)}(s')$ the Born cross section and $R(z,s/m_e^2)$ the distribution--valued radiator 
\cite{DISTR} describing the initial state radiation of photons and light $e^+e^-$ pairs. The lower bound $s_0 = 
s z_0$ is an invariant mass squared depending on the experiment, cf.~\cite{ALEPH:2005ab}. In the later numerical 
illustrations we will choose $s_0 = 4 m_\tau^2$, with $m_\tau$ the $\tau$ lepton mass.

The general decomposition of the scattering cross section in Mellin space is given by, cf.~\cite{Berends:1987ab}\footnote{
In the massless case the principle solution of the RGEs to general 
orders has been known for long, see~\cite{Blumlein:1997em,Ellis:1993rb}.}
%--------------------------------------------------------------------------------------------------------
\begin{eqnarray}
\label{EQ:SI2}
\frac{d\sigma_{e^+e^-}}{ds'} &=& \frac{1}{s} \sigma^{(0)}(s') 
\Biggl[ \Gamma_{e^+e^+}\left(N,\frac{\mu^2}{m_e^2}\right)
\tilde{\sigma}_{e^+e^-}\left(N,\frac{s'}{\mu^2}\right)
\Gamma_{e^-e^-}\left(N,\frac{\mu^2}{m_e^2}\right)
\nonumber\\
&& +
\Gamma_{\gamma e^+}\left(N,\frac{\mu^2}{m_e^2}\right)
\tilde{\sigma}_{e^- \gamma}\left(N,\frac{s'}{\mu^2}\right)
\Gamma_{e^-e^-}\left(N,\frac{\mu^2}{m_e^2}\right)
\nonumber\\
&& +
\Gamma_{e^+ e^+}\left(N,\frac{\mu^2}{m_e^2}\right)
\tilde{\sigma}_{e^+ \gamma}\left(N,\frac{s'}{\mu^2}\right)
\Gamma_{\gamma e^-}\left(N,\frac{\mu^2}{m_e^2}\right)
\nonumber\\
&& +
\Gamma_{\gamma e^+}\left(N,\frac{\mu^2}{m_e^2}\right)
\tilde{\sigma}_{\gamma \gamma}\left(N,\frac{s'}{\mu^2}\right)
\Gamma_{\gamma e^-}\left(N,\frac{\mu^2}{m_e^2}\right)
\Biggr].
\end{eqnarray}
%--------------------------------------------------------------------------------------------------------
The terms in the brackets $[...]$ are Mellin--convoluted. Only massive OMEs of the kind 
$\Gamma_{e^{\pm} e^{\pm}}$ and $\Gamma_{\gamma e^{\pm}}$ 
contribute because the process considered has electron--positron initial states.
The last term in Eq.~(\ref{EQ:SI2}) is only contributing with $O(a^4)$. $\mu$ denotes the 
factorization and renormalization scale. As we will see later, it will cancel in the scattering cross section,
when performing the expansion consistently to a certain order in $a_0$.
The massive OMEs, $\Gamma_{ij}$, and massless Wilson coefficients, $\tilde{\sigma}_{ij}$, obey the following 
series representations
%--------------------------------------------------------------------------------------------------------
\begin{eqnarray}
\Gamma_{li}\left(N, \frac{\mu^2}{m_e^2}\right)       &=& \delta_{li} + \sum_{r=1}^\infty a^r(\mu^2) \sum_{n=0}^r 
a_{li;nr}(N) \Lambda^n
\\
\tilde{\sigma}_{lk}\left(N, \frac{s'}{\mu^2} \right) &=& \delta_{lk} + \sum_{r=1}^\infty a^r(\mu^2) \sum_{n=0}^r 
b_{lk;nr}(N) 
\lambda^n,
\end{eqnarray}
%--------------------------------------------------------------------------------------------------------
with the logarithms $\Lambda$ and $\lambda$ given by
%--------------------------------------------------------------------------------------------------------
\begin{eqnarray}
\Lambda        = \ln\left(\frac{\mu^2}{m_e^2}\right),~~~~~~~
\lambda  = \ln\left(\frac{s'}{\mu^2}\right).
\end{eqnarray}
%--------------------------------------------------------------------------------------------------------
The massive OMEs $\Gamma_{ij}$ and massless Wilson coefficients $\tilde{\sigma}_{kl}$ fulfill the 
following renormalization group equations, cf.~\cite{Blumlein:2000wh}, 
%--------------------------------------------------------------------------------------------------------
\begin{eqnarray}
\label{eq:RE1}
\left[\left(\frac{\partial}{\partial \Lambda} + \beta(a) \frac{\partial}{\partial a}\right) \delta_{kl} + 
\frac{1}{2} 
\gamma_{kl}(N,a)\right]
\Gamma_{li}\left(N,a,\frac{\mu^2}{m_e^2}\right) 
&=& 0 \\
%----
\label{eq:RE2}
\left[\left(\frac{\partial}{\partial \lambda} - \beta(a) \frac{\partial}{\partial a}\right) \delta_{kl} \delta_{jm} 
+ \frac{1}{2} \gamma_{kl}(N,a) \delta_{jm}
+ \frac{1}{2} \gamma_{jm}(N,a) \delta_{kl} \right]
\tilde{\sigma}_{lj}\left(N,a,\frac{s'}{\mu^2}\right) &=& 0~, 
\end{eqnarray}
%--------------------------------------------------------------------------------------------------------
and the QED $\beta$ function has the representation 
%--------------------------------------------------------------------------------------------------------
\begin{eqnarray}
\beta(a) = - \sum_{k=0}^\infty \beta_k a^{k+2}.
\end{eqnarray}
%--------------------------------------------------------------------------------------------------------
Eq.~(\ref{eq:12}) gives an overview on the orders of the expansion of the radiators in the fine 
structure constant which are now available, including the results of the present calculation, 
%------------------------------------------------------------------------------------------------------------
\begin{equation}
\label{eq:12}
\begin{array}{lll}
\alpha \ {L}   &  \alpha                 &       
\\
\alpha^2 { L}^2 &  \alpha^2 {L}   & \alpha^2
\\
\alpha^3 {L}^3 &  \alpha^3 {L}^2 & \alpha^3 {L}
\\
\alpha^4 {L}^4 &  \alpha^4 {L}^3 & \alpha^4 {L}^2
\\
\alpha^5 {L}^5 &  \alpha^5 {L}^4 & \alpha^5 {L}^3
\\
\alpha^6 {L}^6 &  \alpha^6 {L}^5~.   &
\end{array}
\end{equation}
%------------------------------------------------------------------------------------------------------------
The expansion coefficients $c_{k, l}$ of Eq.~(\ref{EQ:SI1}) up to the sixth order in $a_0$ in Mellin space are 
given by Eqs.~(\ref{EQ:EXP1}--\ref{EQ:EXP21}). They are expressed by the anomalous dimensions 
$\gamma_{ij}^{(k)}$ 
\cite{NLOand}, the expansion
coefficients of the massless Drell--Yan cross section \cite{Hamberg:1990np,Harlander:2002wh,Blumlein:2019srk},
$\sigm_{ij}^{(k)}$, the expansion coefficients of the QED $\beta$ function and the renormalized massive OMEs 
$\Gamma_{ij}^{(k)}$, where $(k+1)$ denotes the
loop order. In the following we use the notation
%--------------------------------------------------------------------------------------------------------
\begin{eqnarray}
 \gamma_{ij}^{(k)} = - P_{ij}^{(k)}(N) = - \Mvec[P_{ij}^{(k)}(z)](N).
\end{eqnarray}
%--------------------------------------------------------------------------------------------------------
When working in $z$ space we will use $P_{ij}^{(k)}(z)$ instead of the anomalous dimensions 
$\gamma_{ij}^{(k)}$.

One obtains
%--------------------------------------------------------------------------------------------------------
\input{formu2.tex}
%--------------------------------------------------------------------------------------------------------
where $P_{\gamma\gamma}^{(0)} = - 2 \beta_0$.
These functions are derived by solving the renormalization group equations for the massive OMEs and 
the massless Wilson coefficients, (\ref{eq:RE1}, \ref{eq:RE2}) up to six--loop order. We show also the
constant term $c_{3,0}$ which contains the term $\Gamma_{ee}^{(2)}$, not having been calculated yet.\footnote{
For QCD the Drell--Yan cross section $\tilde{\sigma}^{(2)}$ has been  calculated 
giving numerical illustrations in \cite{Duhr:2020seh} very recently.} 

A few remarks are in order. The sub--system cross section for the Drell--Yan process is flavor dependent, 
cf.~Eq.~(\ref{EQ:SI2}), which results in the present case from the to $e^+ e^-$ initial state, the vertex to which the 
produced neutral current gauge bosons, $\gamma^*$
or $Z^*$, couple. I.e. e.g. in the term describing intermediate photon exchange in 
Eq.~(\ref{eq:FF2}) by $P_{e\gamma}^{(0)} P_{\gamma e}^{(0)}$ has to be read as either 
$P_{e^-\gamma}^{(0)} P_{\gamma e^-}^{(0)}$ or 
$P_{e^+\gamma}^{(0)} P_{\gamma e^+}^{(0)}$. Therefore, one has 
%--------------------------------------------------------------------------------------------------------
\begin{eqnarray}
P_{e^- \gamma}^{(0)}(z) = P_{e^+ \gamma}^{(0)}(z) = 4 [z^2 + (1-z)^2], 
\end{eqnarray}
%--------------------------------------------------------------------------------------------------------
and the energy--momentum sum--rule 
%--------------------------------------------------------------------------------------------------------
\begin{eqnarray}
\label{eq:sr1}
\int_0^1 dz~z~[P_{ee}^{(0)}(z) + P_{e^- \gamma}^{(0)}(z) + P_{e^+ \gamma}^{(0)}(z)] = 0,
\end{eqnarray}
%--------------------------------------------------------------------------------------------------------
is obeyed. Similar relations hold in higher order and also apply to the corresponding massive OMEs.
$P_{ee}^{(0)}$ is flavor conserving, i.e. it does either describe an $e^- \rightarrow e^-$ or an $e^+ \rightarrow e^+$
transition.

Furthermore, one has
%--------------------------------------------------------------------------------------------------------
\begin{eqnarray}
\label{eq:sr2}
\int_0^1 dz~z~[P_{\gamma e}^{(0)}(z) + P_{\gamma \gamma}^{(0)}(z)] = 0,
\end{eqnarray}
%--------------------------------------------------------------------------------------------------------
where $P_{\gamma e}^{(0)}(z) = P_{\gamma e^+}^{(0)}(z) + P_{\gamma e^-}^{(0)}(z)$, like in Quantum Chromodynamics
(QCD) setting the color factor $C_F = 1$, cf.~\cite{Blumlein:2012bf}. This complication is usually not present in QCD, 
since there the splitting functions act in the singlet case on fully symmetrized flavor distributions, like the flavor 
singlet distribution $\Sigma(x,Q^2)$, summing over all quark and antiquark flavors. 

In the above, $P_{ij}^{(k)}$ denotes the Mellin transform of the  corresponding expansion coefficient  of the 
splitting function. The building blocks for the above quantities are given in Eqs.~(80--82, 90--92, 94, 95) of 
Ref.~\cite{Blumlein:2011mi} and Refs.~\cite{Hamberg:1990np,Harlander:2002wh,Blumlein:2019srk,NLOand}
in $z$ space and the operator matrix element calculated in Section~\ref{sec:3}.   Furthermore, a series of 
Mellin convolutions is needed which are given in Appendix~A of 
\cite{Blumlein:2011mi}. Further higher order convolutions can be carried out by the algorithms encoded in
the package {\tt HarmonicSums}~\cite{Vermaseren:1998uu,Blumlein:1998if,Ablinger:2014rba,Ablinger:2010kw, Ablinger:2013hcp, 
Ablinger:2011te,
Ablinger:2013cf,Ablinger:2014bra,Ablinger:2017Mellin}.

The running coupling constant is the solution of the differential equation
%--------------------------------------------------------------------------------------------------------
\begin{eqnarray}
\label{EQ:run1}
\frac{d a(\mu^2)}{d\ln(\mu^2)} = - \sum_{k=0}^\infty \beta_k a^{k+2}(\mu^2),
\end{eqnarray}
%--------------------------------------------------------------------------------------------------------
with $\beta_k$ the expansion coefficients of the $\beta$-function, with 
%--------------------------------------------------------------------------------------------------------
\begin{eqnarray}
\label{EQ:runb}
\beta_0 = -\frac{4}{3},~~~~~\beta_1 = -4,~~~~~\beta_2 = \frac{62}{9}, 
\end{eqnarray}
%--------------------------------------------------------------------------------------------------------
\cite{Gross:1973id,Politzer:1973fx,Caswell:1974gg,Vladimirov:1979zm} in the single fermion approach, $N_F = 1$, 
retaining only electrons. Since we are dealing with the first three logarithmic orders from $O(\alpha^3)$ onward
only terms up to $\beta_2$ are contributing. The solution of (\ref{EQ:run1}) has been derived in 
Ref.~\cite{Blumlein:1994kw}, Eqs.~(2, 3),  in the $\overline{\sf MS}$ scheme by keeping all terms up to $\beta_2$ 
in closed form. The corresponding perturbative solution for $a(\mu^2) = a(a_0,L)$ is then given by 
%--------------------------------------------------------------------------------------------------------
\begin{eqnarray}
\label{EQ:run3}
a(\mu^2) = a_0 - a_0^2 L \beta_0  
         + a_0^3 L \left(\beta_0^2 L   - \beta_1\right) 
         - a_0^4 L \left(\beta_0^3 L^2 - \frac{5}{2} \beta_0 \beta_1 L + \beta_2\right) + O\left(a_0^5\right). 
\end{eqnarray}
%--------------------------------------------------------------------------------------------------------
One verifies the correctness of (\ref{EQ:run3}) by inserting it into Eq.~(\ref{EQ:run1}).
Note that if one does not expand the fine structure constant $a(\mu^2)$ w.r.t. its reference value $a_0 = a(\mu^2 = 
m_e^2)$, the $\mu$ dependence in the RGE--decomposition given in Ref.~\cite{Blumlein:2011mi} is not canceled.
However, expanding to the respective order in the coupling constant intended, a scheme--invariant expression is
obtained. This is the reason, why we are expanding the logarithmic dependence of the coupling constant and 
write the scattering cross section in terms of $a_0$.

In the above expressions we presented the sub--system cross sections $\tilde{\sigma}_{ij}$ in a genuine way. One should
note that these quantities are partly different for vector and axial--vector couplings, cf.~\cite{Hamberg:1990np,QED2019}
and so are some of the radiators. The first expressions for $c_{i,j}$ in $z$ space are presented in 
Section~\ref{sec:4}. In Appendix~\ref{sec:B} all coefficients $c_{i,j}$ in Mellin $N$ space, which are more compact, 
are given.

Sumrules do not only hold for the splitting functions (\ref{eq:sr1}--\ref{eq:sr2}) but also for the universal 
unrenormalized massive operator matrix elements, cf.~\cite{Buza:1996wv,BBK2} to two--loop order
%--------------------------------------------------------------------------------------------------------
\begin{eqnarray}
\int_0^1 dx  x\Biggl[
\hat{A}_{ee}^{\rm NS}  + \hat{A}_{ee}^{\rm PS} + \hat{A}_{\gamma e} \Biggr] &=& 1,
%\\
%\int_0^1 dx \Biggl[
%A_{e\gamma}^{\rm NS}  + A_{\gamma \gamma} \Biggr] &=& 1
\end{eqnarray}
%--------------------------------------------------------------------------------------------------------
which we have verified for $N_F = 1$.
%--------------------------------------------------------------------------------------------------------
\section{The Operator Matrix Element \boldmath $\Gamma^{(1)}_{\gamma e}$}
\label{sec:3}
%--------------------------------------------------------------------------------------------------------

\vspace*{1mm}
\noindent
For the calculation of the operator matrix element $\Gamma^{(1)}_{\gamma e}$ we follow the notation of~\cite{Blumlein:2011mi}.
After wave function and mass renormalization we can write the operator matrix element as
%--------------------------------------------------------------------------------------------------------
\begin{eqnarray}
      \hat{A}_{\gamma e} &=& \hat{a} \cdot \hat{\hat{A}}_{\gamma e}^{(1)} 
      + \hat{a}^2 \left[ \hat{\hat{A}}_{\gamma e}^{(2)} + Z_\text{CT}^{(2)} \right] + O(\hat{a}^3),
\end{eqnarray}
%--------------------------------------------------------------------------------------------------------
where $Z_\text{CT}^{(2)}$ are the counter term contributions due to mass and wave function renormalization,
$\hat{\hat{A}}_{\gamma e}^{(i)}$ are the unrenormalized operator matrix elements at $i$-loops and $\hat{a}$
is the unrenormalized fine structure constant. 

The renormalized operator matrix elements in the $\text{MOM}$-scheme are given by, cf.~\cite{BBK2},
%--------------------------------------------------------------------------------------------------------
\begin{eqnarray}
      A_{\gamma e}^\text{MOM} &=& a^\text{MOM} \left[ \hat{A}_{\gamma e}^{(1)} + Z_{\gamma e}^{-1,(1)} \right]
      + \left. a^\text{MOM} \right.^2 \Bigl[ \hat{A}_{\gamma e}^{(2)} + Z_{\gamma e}^{-1,(2)} +  Z_{\gamma e}^{-1,(1)} \hat{A}_{e e}
\nonumber \\ &&
      + Z_{\gamma \gamma}^{-1,(1)} \hat{A}_{\gamma e} + \delta a_1^\text{MOM} \hat{A}_{\gamma e}^{(1)} \Bigr] + O( \left. a^\text{MOM} \right.^3) ,
\end{eqnarray}
%--------------------------------------------------------------------------------------------------------
where the unrenormalized coupling constant can be expressed via
%--------------------------------------------------------------------------------------------------------
\begin{eqnarray}
      \hat{a} &=& a^\text{MOM} \left[ 1 + \delta a_1^\text{MOM} a^\text{MOM} \right] + O\left(\left. a^\text{MOM} 
\right.^3\right)
\end{eqnarray}
%--------------------------------------------------------------------------------------------------------
with
%--------------------------------------------------------------------------------------------------------
\begin{eqnarray}
      \delta a_1^\text{MOM} = S_\ep \frac{2 \beta_0}{\ep} \left( \frac{m^2}{\mu^2} \right)^{\ep/2} 
      \exp\left[ \sum_{i=0}^\infty \frac{\zeta_i}{i} \left( \frac{\ep}{2} \right)^i \right],
\end{eqnarray}
%--------------------------------------------------------------------------------------------------------
and the spherical factor $S_\ep = \exp[(\ep/2)[\ln(4\pi) - \gamma_E]]$.

The operator matrix element, after charge and wave function renormalization, can therefore be written as
%--------------------------------------------------------------------------------------------------------
\begin{eqnarray}
      \hat{A}_{\gamma e} &=& a^\text{MOM} S_\ep \left( \frac{m^2}{\mu^2} \right)^{\ep/2} \left[ -\frac{1}{\ep} P_{\gamma e}^{(0)} + \Gamma_{\gamma e}^{(0)} + \ep \bar{\Gamma}_{\gamma e}^{(0)} \right]
      + \left. a^\text{MOM} \right.^2 S_\ep^2 \left( \frac{m^2}{\mu^2} \right)^{\ep} \Biggl\{
         \frac{\gamma_{\gamma e}^{(0)}}{2\ep^2} 
         \bigl[
             \gamma_{ee}^{(0)} + \gamma_{\gamma\gamma}^{(0)} - 4 \beta_0 
         \bigr] 
\nonumber \\ &&
       + \frac{1}{\ep} 
         \biggl[
            \frac{1}{2} \gamma_{\gamma e}^{(1)} 
            + \gamma_{\gamma e}^{(0)} \Gamma_{ee}^{(1)}
            + \gamma_{\gamma \gamma}^{(0)} \Gamma_{\gamma e}^{(1)}
            - 2 \beta_0 \Gamma_{\gamma e}^{(1)}
         \biggr] 
       + \hat{\Gamma}_{\gamma e}^{(1)} \Biggr\}~.
\end{eqnarray}
%--------------------------------------------------------------------------------------------------------
The predicted pole structure serves as a test on the calculation.
The renormalized OME in the $\MS$-scheme is given by
%--------------------------------------------------------------------------------------------------------
\begin{eqnarray}
      A_{\gamma e} &=& a^{\MS} \left[ - \frac{1}{2} P_{\gamma e}^{(0)} L + \Gamma_{\gamma e}^{(0)} \right]
      + \left. a^{\MS} \right.^2 \Bigl[ 
      \frac{P_{\gamma e}^{(0)}}{8} \left( P_{ee}^{(0)} + P_{\gamma\gamma}^{(0)} + 2 \beta_0 \right) L^2 
\nonumber \\ &&
      + \frac{1}{2} \left( P_{\gamma e}^{(1)} 
      + \Gamma_{e e}^{(0)} P_{\gamma e}^{(0)}  
      + \Gamma_{\gamma e}^{(0)} P_{\gamma\gamma}^{(0)} 
      + 2 \beta_0 \Gamma_{\gamma e}^{(0)}
      \right) L 
      + \hat{\Gamma}_{\gamma e}^{(1)}
\nonumber \\ &&
      + \bar{\Gamma}_{ee}^{(0)} P_{\gamma e}^{(0)}
      + \bar{\Gamma}_{\gamma e}^{(0)} P_{\gamma \gamma}^{(0)}
      + 2 \beta_0 \bar{\Gamma}_{\gamma e}^{(0)} 
      \Bigr],
\end{eqnarray}
%--------------------------------------------------------------------------------------------------------
where we used the relation
%--------------------------------------------------------------------------------------------------------
\begin{eqnarray}
      a^{\text{MOM}} = a^{\MS} + \beta_0 L \left. a^{\MS} \right.^2 
\end{eqnarray} 
%--------------------------------------------------------------------------------------------------------
and 
%--------------------------------------------------------------------------------------------------------
\begin{eqnarray}
\hat{\Gamma}_{ij} = {\Gamma}_{ij}(N_F+1) - {\Gamma}_{ij}(N_F).  
\end{eqnarray} 
%--------------------------------------------------------------------------------------------------------

\begin{figure}[H]
\centering

\vspace*{-1cm}
\includegraphics[width=0.19\linewidth,height=20cm]{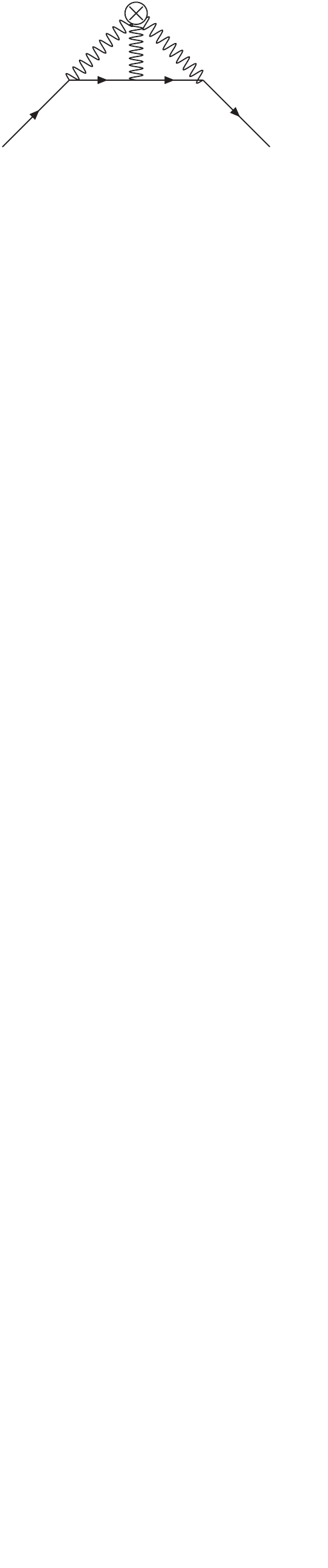}
\includegraphics[width=0.19\linewidth,height=20cm]{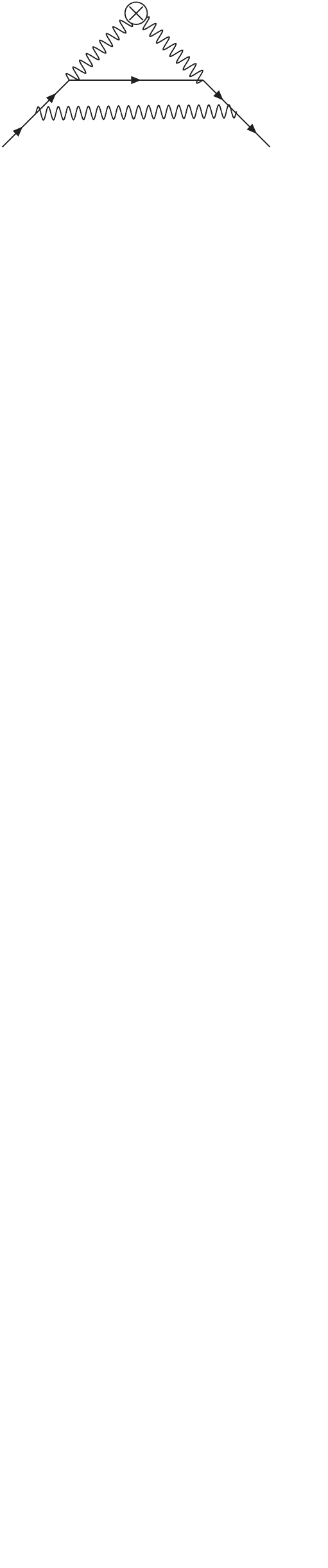}
\includegraphics[width=0.19\linewidth,height=20cm]{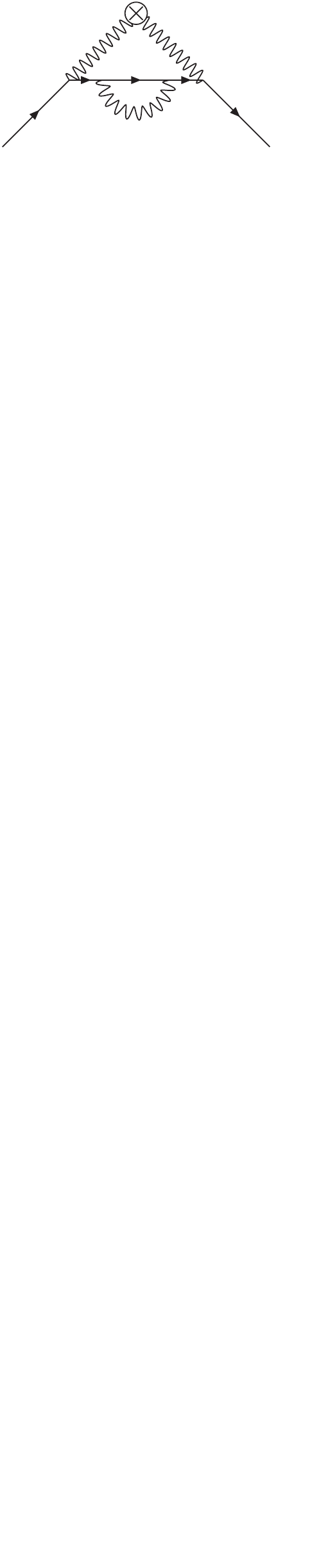}
\includegraphics[width=0.19\linewidth,height=20cm]{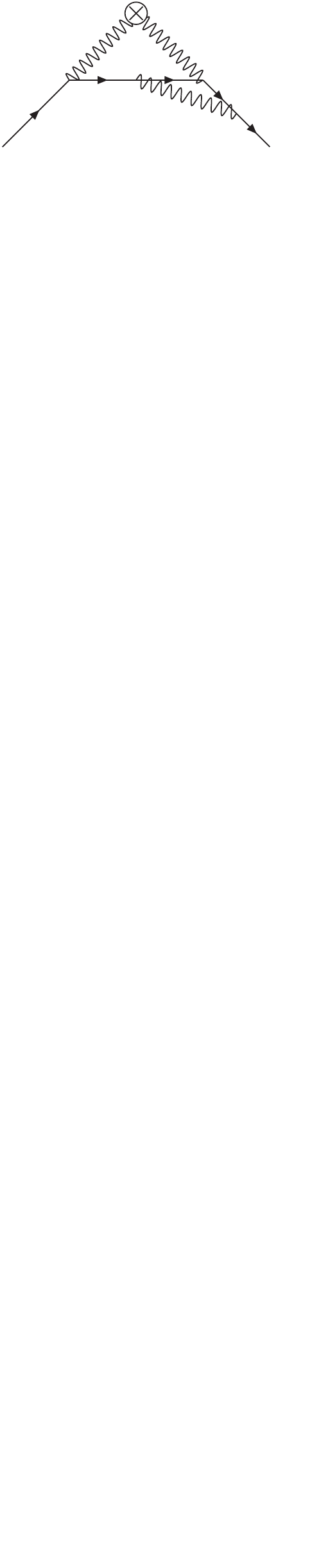}
\includegraphics[width=0.19\linewidth,height=20cm]{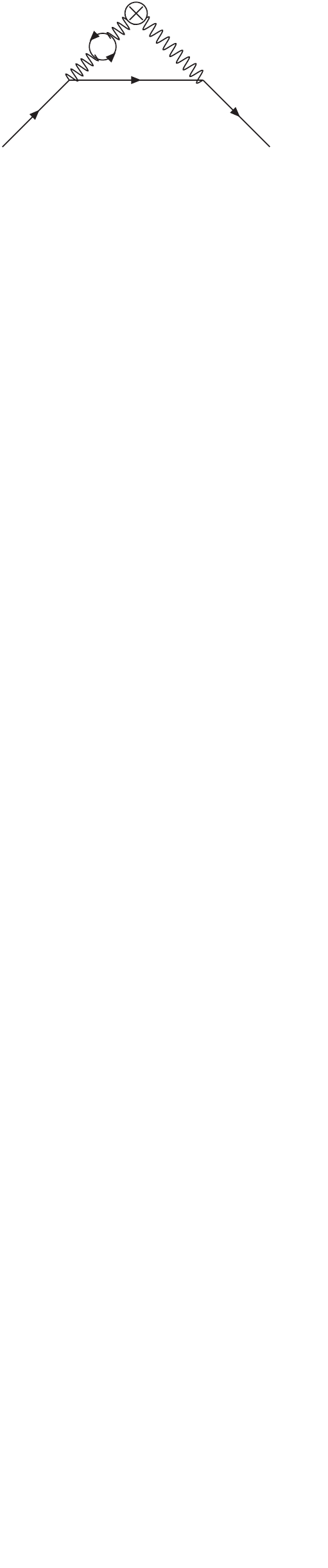}

\vspace*{-17cm}
\includegraphics[width=0.19\linewidth,height=20cm]{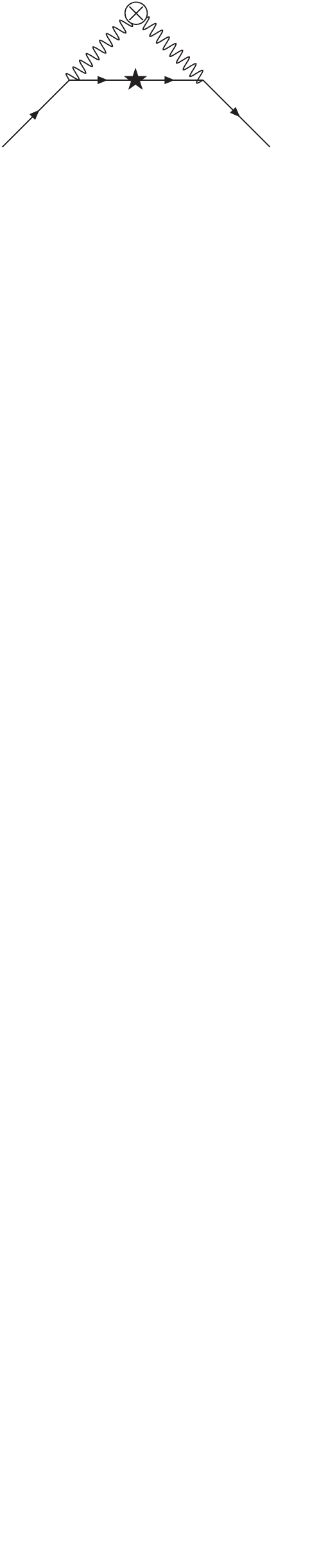}
\includegraphics[width=0.19\linewidth,height=20cm]{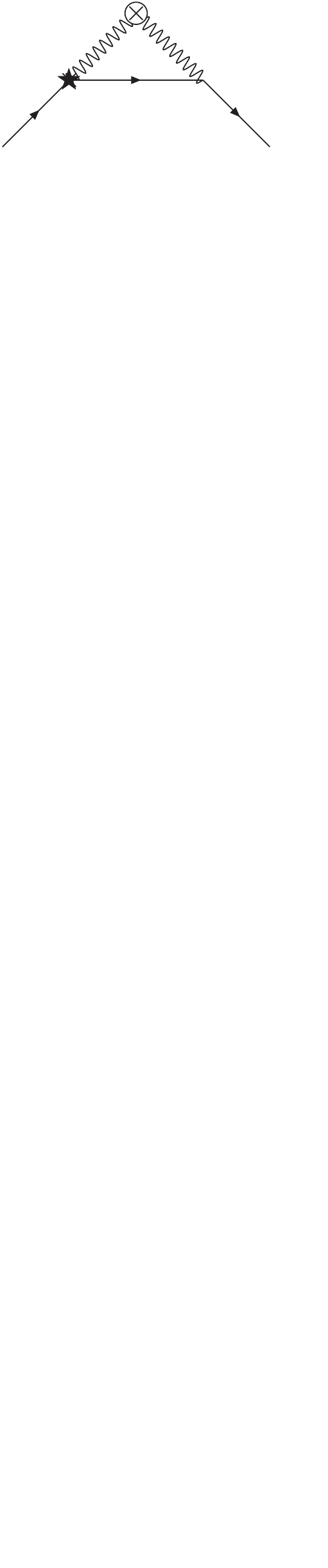}

\vspace*{-17cm}
\caption{\sf
The Feynman diagrams contributing to $\Gamma^{(1)}_{\gamma e}$. For the notation and the Feynman rules see
Ref.~\cite{Blumlein:2011mi}. The symbol {\tiny $\bigstar$} denotes the counter-term insertion.}
\label{DIAG1}
\end{figure}

The Feynman diagrams contributing to $\Gamma^{(1)}_{\gamma e}$ are shown in Figure~\ref{DIAG1}, with the 
corresponding 
symmetrization understood. They can be represented in terms of 18 master integrals by performing the 
integration-by-parts 
reduction using the package {\tt Litered} \cite{LR}. Here, as in 
previous calculations, 
cf.~\cite{Ablinger:2014vwa}, we resummed the local operators into linear 
propagators. The master integrals are either calculated using conventional methods like generalized hypergeometric 
functions, cf. e.g. \cite{Blumlein:2018cms}, or can be obtained by solving ordinary differential equations, cf. 
e.g.~\cite{Ablinger:2018zwz}.\footnote{We took the opportunity to re--calculate the results of \cite{Blumlein:2011mi}
by using the same techniques. Here also 18 master integrals contribute, which can be calculated in a similar manner.
This can now be done in a fully automated way. The large mathematical and conceptional progress in performing loop 
integrals since 2002 is clearly demonstrated by this.}
The code {\tt Tarcer} \cite{Mertig:1998vk} has been used for checks and to determine
initial values.
%-----------------------------------------------------------------------------------------------------------------
%-----------------------------------------------------------------------------------------------------------------
\noindent

The expansion coefficients of the unrenormalized OME $\hat{A}_{\gamma e}^{(1)}$ are given by
%-----------------------------------------------------------------------------------------------------------------
\begin{eqnarray}
\hat{A}_{\gamma e}^{(1)} &=& 
\hat{a} \left(\frac{m_e^2}{\mu^2}\right)^{\ep/2} S_\ep \left[
- \frac{1}{\ep} P_{\gamma e}^{(0)}(z) 
+ \Gamma_{\gamma e}^{(0)}(z)
+ \ep \overline{\Gamma}_{\gamma e}^{(0)}(z) + O(\ep^2) \right],
\end{eqnarray}
%-----------------------------------------------------------------------------------------------------------------
see \cite{Blumlein:2011mi}.
The coefficient $\Gamma_{\gamma e}^{(0)}$ reads
%-----------------------------------------------------------------------------------------------------------------
\begin{eqnarray}
\Gamma_{e \gamma}^{(0)} &=& -2 \frac{1 + (1-z)^2}{z}\left[2 \ln(z) +1\right].
\end{eqnarray}
%-----------------------------------------------------------------------------------------------------------------

The corresponding result in Mellin $N$ space is the obtained as the $N$th expansion coefficient in the auxiliary variable 
$\hat{x}$ and the $z$ space representation can be obtained by a subsequent inverse Mellin transform.
The renormalized OME $A_{\gamma e}^{(2)}$ is given by
%-----------------------------------------------------------------------------------------------------------------
\begin{eqnarray}
\label{eq:AN}
      A_{\gamma e}^{(2)}(N) &=&
      \Biggl[ 
            \frac{\big(N^2+N+2\big)\big(N^2+N+6\big)}{3 (N-1) N^2 (N+1)^2}
            -\frac{4 \big(N^2+N+2\big)}{(N-1) N (N+1)} S_1
      \Biggr] L^2
\nonumber \\ &&
      - \Biggl[ 
            \frac{2 P_2}{9 (N-1)^2 N^3 (N+1)^3}
            -\frac{4 P_1}{3 (N-1) N^2 (N+1)^2} S_1
            +\frac{12 \big(N^2+N+2\big)}{(N-1) N (N+1)} S_1^2
\nonumber \\ &&
            +\frac{12 \big(N^2+N+2\big)}{(N-1) N (N+1)} S_2
      \Biggr] L
      +\frac{P_ 8}{27 (N-4) (N-3) (N-2) (N-1) N^4 (N+1)^4}
\nonumber \\ &&
      +\biggl(
            \frac{2 P_ 7}{9 (N-4) (N-3) (N-2) (N-1) N^3 (N+1)^3}
            +\frac{2 \big(N^2+N+2\big)}{(N-1) N (N+1)} S_ 2            
      \biggr) S_ 1
\nonumber \\ &&
      +\frac{P_ 3}{3 (N-2) (N-1) N (N+1)^2} S_ 1^2
      +\frac{2 \big(N^2+N+2\big)}{3 (N-1) N (N+1)} S_ 1^3
\nonumber \\ &&
      +\frac{P_ 6}{3 (N-2) (N-1) N^2 (N+1)^2} S_ 2
      +\frac{4 \big(N^2+N+2\big)}{3 (N-1) N (N+1)} S_ 3
\nonumber \\ &&
      -\frac{48 \big(N^2+N+2\big)}{(N-1) N (N+1)} S_{2,1}
      +\frac{3 \cdot 2^{6+N}}{(N-2) (N+1)^2} S_ {1,1}\biggl(\frac{1}{2},1\biggr)
\nonumber \\ &&
      + \frac{2^{6-N} P_ 5}{3 (N-3) (N-2) (N-1)^2 N^2} 
      \biggl(
              S_ 2(2)
            + S_ 1 S_1(2)
            - S_{1,1}(1,2)
            - S_{1,1}(2,1)
      \biggr)
\nonumber \\ &&
      -\frac{32 \big(N^2+N+2\big)}{(N-1) N (N+1)}
      \biggl[
            S_ 1(2) S_ {1,1}\biggl(\frac{1}{2},1\biggr)
            + S_ {1,2}\biggl(\frac{1}{2},2\biggr)
            - S_ {1,1,1}\biggl(\frac{1}{2},1,2\biggr)
\nonumber \\ &&
            - S_ {1,1,1}\biggl(\frac{1}{2},2,1\biggr)
            - \frac{\zeta_2}{2} S_1(2)
      \biggr]
      + \frac{4 P_ 4}{(N-2) (N-1) N^2 (N+1)^2} \zeta_2,
\end{eqnarray}
with the polynomials
\begin{eqnarray}
      P_1 &=& 25 N^4+44 N^3+87 N^2+56 N+12,
\\
      P_2 &=& 112 N^7+194 N^6+347 N^5+339 N^4+93 N^3-293 N^2-60 N+36,
\\ 
      P_3 &=& 17 N^4-66 N^3-179 N^2-272 N-212,
\\
      P_4 &=& N^5+4 N^4 +25 N^3+14 N^2+12 N+8 -3 \cdot 2^{N+3} N^2 \bigl(N - 1\bigr),
\\
      P_5 &=& 9 N^5-24 N^4+8 N^3+4 N^2+33 N-18,
\\
      P_6 &=& 11 N^5-90 N^4-329 N^3-356 N^2-284 N-48,
\\
      P_7 &=& 17 N^9+213 N^8-1729 N^7+2329 N^6-5196 N^5+7898 N^4+16196 N^3
\nonumber \\ &&
      +12528 N^2-4896 N-3456,
\\
      P_8 &=& -509 N^{11}+2365 N^{10}+2797 N^9-13158 N^8+31274 N^7-4694 N^6-64636 N^5
\nonumber \\ &&
      -107861 N^4-14622 N^3+6588 N^2-2376 N-2592.
\end{eqnarray}
%-----------------------------------------------------------------------------------------------------------------
$A_{\gamma e}^{(2)}$ is expressed by harmonic sums \cite{Vermaseren:1998uu,Blumlein:1998if}
%-----------------------------------------------------------------------------------------------------------------
\begin{eqnarray}
S_{b,\vec{a}}(N) = \sum_{k=1}^N \frac{({\rm sign}(b))^k}{k^{|b|}} S_{\vec{a}}(k),~~~S_\emptyset = 1, a_i,b_i \in 
\mathbb{Z} \backslash \{0\}, 
\end{eqnarray}
%-----------------------------------------------------------------------------------------------------------------
and generalized harmonic sums \cite{Moch:2001zr,Ablinger:2013cf} 
%-----------------------------------------------------------------------------------------------------------------
\begin{eqnarray}
S_{b,\vec{a}}(d,\vec{c},N) = \sum_{k=1}^N \frac{d^k}{k^{|b|}} S_{\vec{a}}(\vec{c},k),~~~S_\emptyset = 1,  
a_i,b \in
\mathbb{N} \backslash \{0\}, c_i,d \in \mathbb{Z} \backslash \{0\},
\end{eqnarray}
%-----------------------------------------------------------------------------------------------------------------
has its rightmost pole at $N=1$ and is otherwise regular. In particular one may show that
%-----------------------------------------------------------------------------------------------------------------
\begin{eqnarray}
A_{\gamma e}^{(2)}(2) &=& 
- \frac{64}{9} L^2
- 32 L
- \frac{149}{81} 
+ \frac{736}{9} \zeta_2  
- 128 \ln(2) \zeta_2 + 32 \zeta_3,
\\
A_{\gamma e}^{(2)}(3) &=& 
- \frac{287}{72} L^2
- \frac{1961}{144} L
- \frac{870277}{10368} 
+ 
\frac{1121}{18} \zeta_2,
\\
A_{\gamma e}^{(2)}(4) &=& 
- \frac{869}{300} L^2
- \frac{87689}{9000} L
- \frac{10336457}{360000}
+ \frac{2027}{75} \zeta_2,
\end{eqnarray}
%-----------------------------------------------------------------------------------------------------------------
applying the algorithms of package {\tt HarmonicSums} 
\cite{Vermaseren:1998uu,Blumlein:1998if,Ablinger:2014rba,Ablinger:2010kw, Ablinger:2013hcp, Ablinger:2011te,
Ablinger:2013cf,Ablinger:2014bra,Ablinger:2017Mellin}.
In $z$ space the OME is given by
%-----------------------------------------------------------------------------------------------------------------
\begin{eqnarray}
\label{eq:Ax}
      A_{\gamma e}^{(2)}(z) &=& 
                \Biggl[
                        -\frac{16-28 z+11 z^2}{3 z}
                        +2 (2-z) \HA_0
                        -\frac{4 \big(2-2 z+z^2\big)}{z}  \HA_1
                \Biggr] L^2
                \nonumber \\ &&
                + \Biggl[
                        -\frac{2 \big(32-5 z+85 z^2\big)}{9 z}
                        +\frac{2 \big(32-32 z+31 z^2\big)}{3 z} \HA_0
                        -2 (2-z) \HA_0^2
                \nonumber \\ &&
                        +\frac{4 \big(20-8 z+13 z^2\big)}{3 z} \HA_1
                        -\frac{12 \big(2-2 z+z^2\big)}{z} \HA_1^2
                        +8 (2-z) \HA_{0,1}
                \nonumber \\ &&
                        -8 (2-z) \zeta_2
                \Biggr] L
                +\frac{P_9}{135 z^3} 
                -\frac{320-335 z+231 z^2}{15 z} \HA_0
                +\frac{12+23 z}{6} \HA_0^2
                +\frac{2-z}{3} \HA_0^3
                \nonumber \\ &&
                +32 (2-z) \bigg(
                        \frac{(2-z)^2}{3 z^2}
                        - \HA_0
                \bigg) \big( \HAA_{-1} \HAA_0 - \HAA_{0,-1} \big)
                -8 (2-z) \HA_{0,0,1}
                \nonumber \\ &&
                -\frac{96-190 z+118 z^2-41 z^3}{3 z^2} \HA_1^2
                -32 (2-z) \big(
                          \HAA_{-1} \HAA_ 0
                        - \HAA_{0,-1}
                \big) \HAA_1
                \nonumber \\ &&
                -\bigg(
                        \frac{2 \big(32-48 z+36 z^2-13 z^3\big)}{3 z^2}
                        +4 (2-z) \HA_0
                \bigg) \HA_{0,1}
                -\biggl(
                         \frac{2 P_{10}}{45 z^4} 
                \nonumber \\ &&
                        -\frac{2 \big(32-48 z+12 z^2+7 z^3\big)}{3 z^2} \HA_0
                \biggr) \HA_1
                + \frac{ 2 \big(2-2 z+z^2\big)}{z}
                \biggl(
                        \frac{\HA_1^3}{3}
                        + 8 \HA_1 \HA_{0,1}
                \nonumber \\ &&
                        + 16 \HAA_0 \HAA_{0,-1}
                        - 32 \HAA_{0,0,-1}
                        - 16 \HA_{0,1,1}
                        + 8 \HAA_0 \zeta_2
                \biggr)
                +\biggl(
                         \frac{4 \big(32-48 z+24 z^2-3 z^3\big)}{3 z^2}
                \nonumber \\ &&
                        -8 (2-z) \big( \HA_0 + 2 \HAA_1 \big) 
                \biggr) \zeta_2
                +\frac{8 \big(12-10 z+5 z^2\big) }{z} \zeta_3,
\end{eqnarray}
%-----------------------------------------------------------------------------------------------------------------
with
%-----------------------------------------------------------------------------------------------------------------
\begin{eqnarray}
        P_9 &=& 1536-3072 z+1312 z^2-316 z^3-2005 z^4 ,
        \\
        P_{10} &=& 256-640 z-400 z^2+1320 z^3-1440 z^4+819 z^5 .
\end{eqnarray}
%-----------------------------------------------------------------------------------------------------------------
Here we refer to the harmonic polylogarithms \cite{Remiddi:1999ew},
%-----------------------------------------------------------------------------------------------------------------
\begin{eqnarray}
&&\HA_{b,\vec{a}}(z) = \int_0^z dz f_b(z) \HA_{\vec{a}}(z),~~\HA_\emptyset = 1,~~f_0(z) = \frac{1}{z},~~f_1(z) = 
\frac{1}{1-z},~~f_{-1}(z) = \frac{1}{1+z},
\nonumber\\ 
&& \HA_{\underbrace{\mbox{\scriptsize 0 \ldots 0}}_k}(z) = \frac{1}{k!} \ln^k(z),
\end{eqnarray}
%-----------------------------------------------------------------------------------------------------------------
and define
%-----------------------------------------------------------------------------------------------------------------
\begin{eqnarray}
{\HA}^*_{\vec{a}}(z) := \HA_{\vec{a}}(1-z).
\end{eqnarray}
%-----------------------------------------------------------------------------------------------------------------

There are some terms in Eq.~(\ref{eq:Ax}) which are $\propto 1/z^4$. This is a reflection of terms $\propto 
1/(N-4)$ in Eq.~(\ref{eq:AN}). One derives the small $z$ expansion of $A_{\gamma e}^{(2)}(z)$ which is 
given by
%-----------------------------------------------------------------------------------------------------------------
\begin{eqnarray}
A_{\gamma e}^{(2)}(z) &=& 
\biggl[
        - \frac{16}{3 z}
        +4/3  
        + 4 \HA_0
\biggr] L^2
+ \biggl[
        - \frac{64}{9z} 
        + \frac{64}{3z} \HA_0
        + \frac{250}{9} 
        - 16 \zeta_2 
        - \frac{64}{3} \HA_0 
        - 4 \HA_0^2
\biggr] L
\nonumber\\ &&
+\frac{448}{27 z} 
-\frac{796}{27} 
+ 16 \zeta_3 
+ (1 + 16 \zeta_2) \HA_0 + 2 \HA_0^2
 + \frac{2}{3} \HA_0^3 + O(z),
\end{eqnarray}
%-----------------------------------------------------------------------------------------------------------------
showing that the most singular terms are of $O(1/z)$.

%--------------------------------------------------------------------------------------------------------
\section{The Radiators in \boldmath $z$ Space}
\label{sec:4}
%--------------------------------------------------------------------------------------------------------

\vspace*{1mm}
\noindent
We express the expansion coefficients (\ref{EQ:EXP1}--\ref{EQ:EXP2}) in terms of harmonic polylogarithms 
\cite{Remiddi:1999ew}.
The corresponding expressions can be obtained using the package {\tt HarmonicSums} \cite{Vermaseren:1998uu,
Blumlein:1998if,Ablinger:2014rba,Ablinger:2010kw, Ablinger:2013hcp, Ablinger:2011te,Ablinger:2013cf,
Ablinger:2014bra,Ablinger:2017Mellin}, after reducing the algebraic relations \cite{Blumlein:2003gb}.
Because of the occurrence of some denominators $1/(1 \pm z)^l,~~l \in \mathbb{N},~~l > 1,$ one performs 
the corresponding series expansion and uses summation techniques encoded in the packages {\tt Sigma} 
\cite{SIG1,SIG2}, {\tt EvaluateMultiSums} and {\tt SumProduction} \cite{EMSSP}. 
As usual, one has to separate the radiators into the part $\propto \delta(1-z)$, the contribution due 
to $+$-distributions and the regular part,
%--------------------------------------------------------------------------------------------------------
\begin{eqnarray}
R(z,{{L}}) = R_\delta({{L}}) \delta(1-z) 
+ [R_+(z,{{L}})]_+  
+ R_{\rm reg}(z,{{L}}).  
\end{eqnarray}
%--------------------------------------------------------------------------------------------------------
For the inclusive cross section the integral (\ref{eq:incl}) has to account for the fact that the 
radiator $R(z,{{L}})$ is distribution--valued \cite{DISTR}.

For the inverse Mellin transform we use the notion {\tt PlusFunctionDefinition $\rightarrow$ 2} of the package {\tt 
HarmonicSums},
%--------------------------------------------------------------------------------------------------------
\begin{eqnarray}
\Mvec\left[\frac{\HA_{m_1,\vec{m}}(z) \HA_1^k(z)}{1-z}\right](N) &=& \int_0^1 dz z^{N-1} \left[\HA_{m_1,\vec{m}}(z) - 
\HA_{m_1,\vec{m}}(1) \right] 
\frac{\HA_1^k(z)}{1-z},~~~~m_1 \neq 1,
\\  
\Mvec\left[\frac{\HA_1^k(z)}{1-z}\right](N) &=& \int_0^1 dz \left(z^{N-1} - 1 \right) 
\frac{\HA_1^k(z)}{1-z}.
\end{eqnarray}
%--------------------------------------------------------------------------------------------------------
We use the following notation
%--------------------------------------------------------------------------------------------------------
\begin{eqnarray}
{\cal D}_k(z) = \left(\frac{\ln^k(1-z)}{1-z}\right)_+,~~~~k \in \mathbb{N},~~\text{with}~~\HA_1(z) = - \ln(1-z).
\end{eqnarray}
%--------------------------------------------------------------------------------------------------------
The splitting function $P_{ee}^{(0)}$ is thus given by
%--------------------------------------------------------------------------------------------------------
\begin{eqnarray}
P_{ee}^{(0)}(z) &=&  8 {\cal D}_0(z) - 4 (1+z) + 6 \delta(1-z).
\end{eqnarray}
%--------------------------------------------------------------------------------------------------------
We use the following  conventions
%------------------------------------------------------------------------------------------------------------------------
\begin{eqnarray}
\HA_{\vec{a}}(x) \equiv \HA_{\vec{a}},~~~~~~~~~\HA_{\vec{a}}(1-x) \equiv \HA_{\vec{a}}^*,
\end{eqnarray}
%------------------------------------------------------------------------------------------------------------------------
The expressions up to $O(a^2)$ are given in Ref.~\cite{QED2019}. In the following we distinguish 
the expansion coefficients in the vector and axial--vector case. Coefficients $c_{ij}$ without an index
$v$ or $a$ apply to both cases. We also display the difference terms
%--------------------------------------------------------------------------------------------------------
\begin{eqnarray}
\label{eq:DIFF}
c_{i,j}^{\rm \Delta} = 
  c_{i,j}^{\rm a}
- c_{i,j}^{\rm v}.
\end{eqnarray}
%--------------------------------------------------------------------------------------------------------
The $\delta$--terms are
%--------------------------------------------------------------------------------------------------------
\input{xspaceD.tex}

%--------------------------------------------------------------------------------------------------------
\noindent
The $+$-distributions are given by
%--------------------------------------------------------------------------------------------------------
\input{xspaceP.tex}

%--------------------------------------------------------------------------------------------------------
\noindent
and the regular contributions read
%--------------------------------------------------------------------------------------------------------
\input{xspace1.tex}
%--------------------------------------------------------------------------------------------------------
The radiators depend on the polynomials $P_k|_{k=1}^{881}$ which are too voluminous to be displayed, 
like also the radiators $c_{5,5}$ to $c_{6,5}$. They are given in an ancillary file to this paper.
The radiators exhibit evanescent poles $\propto 1/z^4$, which all cancel by performing an expansion 
around $z = 0$ and the leading pole is again $1/z$. 
%--------------------------------------------------------------------------------------------------------
\section{Numerical Results}
\label{sec:5}
%--------------------------------------------------------------------------------------------------------

\vspace*{1mm}
\noindent
In the following we study the effect of the radiators calculated on the $Z$ resonance. We extend previous work 
\cite{Blumlein:2019pqb} to $O(\alpha^2)$ including the higher order corrections up to $O(\alpha^6 L^5)$ accounting 
for the first three logarithmic corrections from $O(\alpha^3)$ to $O(\alpha^5)$. 
%-----------------------------------------------------------------------------------------------------------------
\begin{figure}[H]
  \centering
  \includegraphics[width=0.6\linewidth]{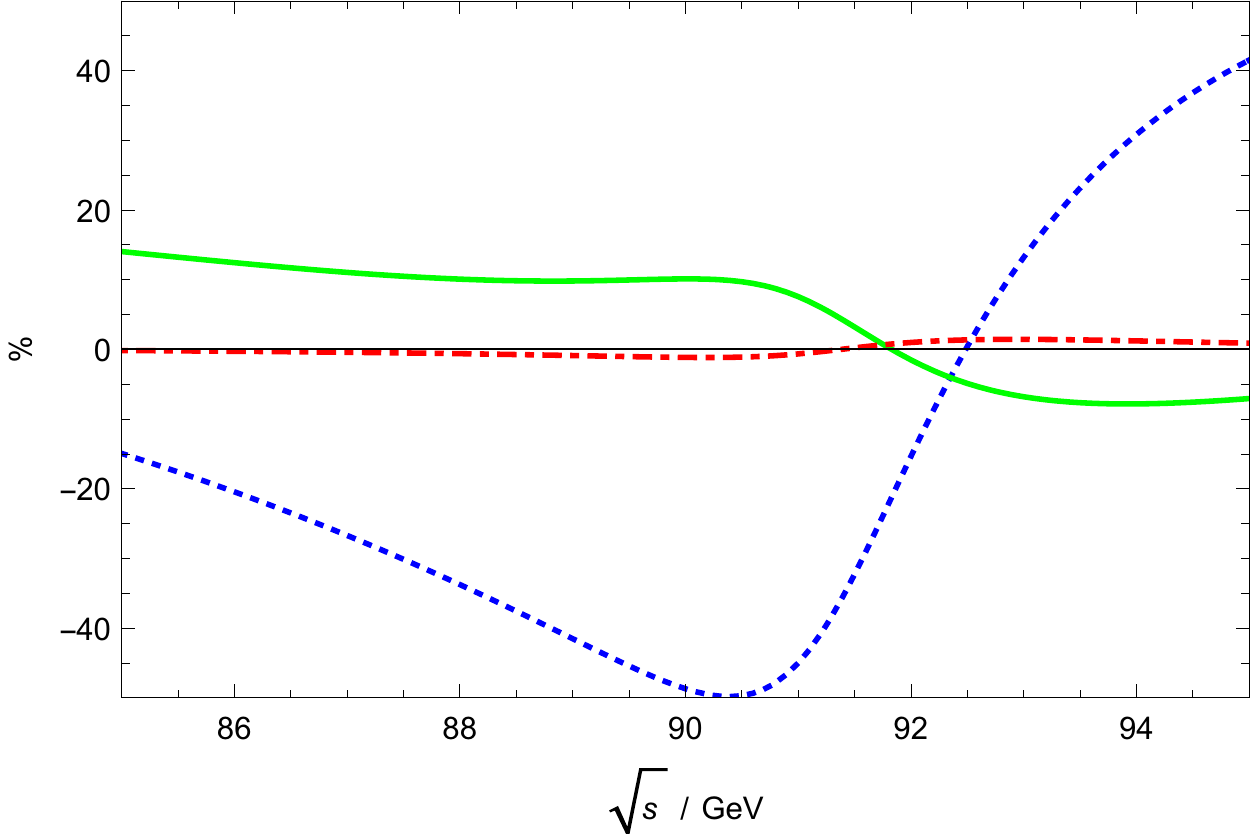}
  \caption{\sf
The ratio of the cross section containing all initial state corrections up 
to $O(a)$ (dotted line),
to $O(a^2)$ (full line),
to $O(a^3 {L}^3)$ (dash-dotted line), 
normalized on all terms including 
also the $O(a^3 {L})$ and $O(a^4 {L}^4)$ corrections to the cross section $e^+ e^- \rightarrow \gamma^*/Z^*$ 
in 
the region of the $Z$-peak
for $M_Z = 91.1876~\GeV$ \cite{PDG}.}
\label{FIG1}
\end{figure}
%-----------------------------------------------------------------------------------------------------------------
%-----------------------------------------------------------------------------------------------------------------
\begin{figure}[H]
  \centering
  \includegraphics[width=0.6\linewidth]{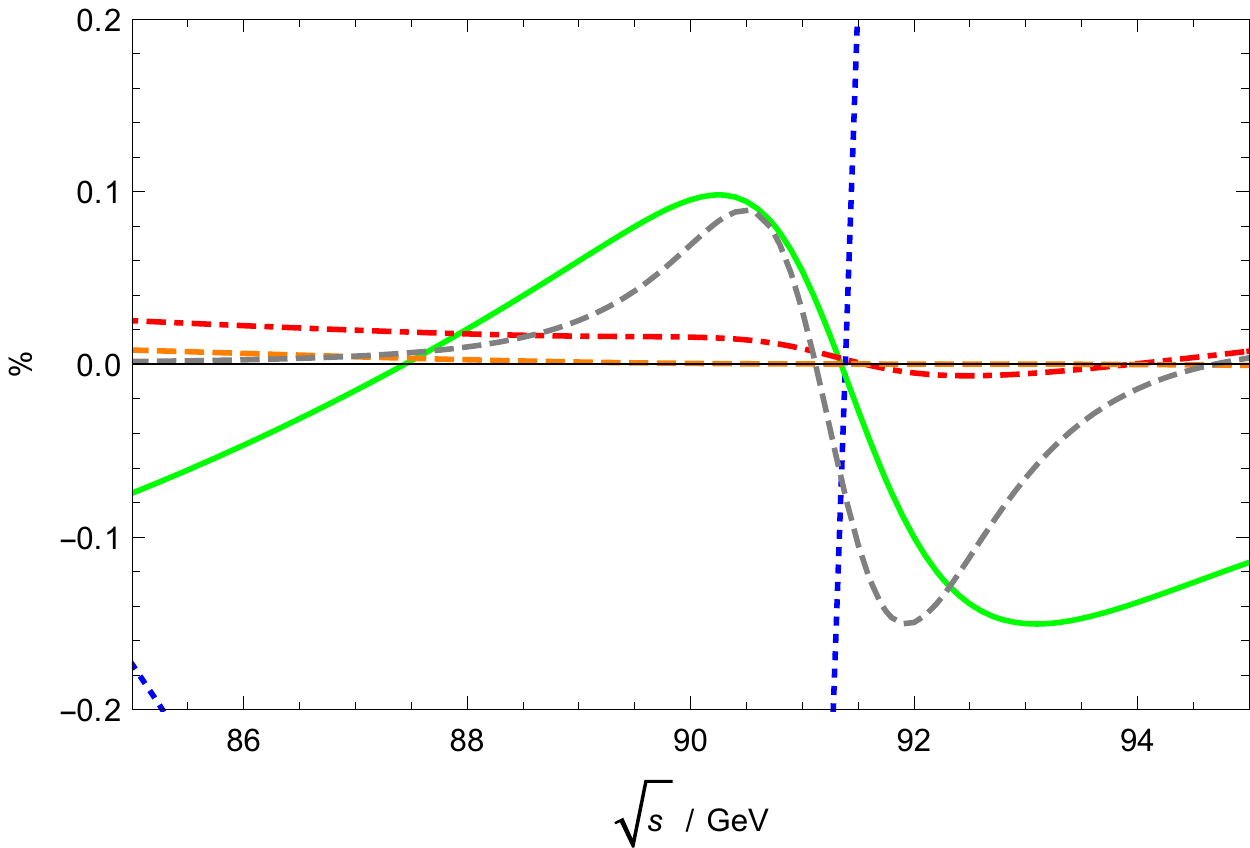}
  \caption{\sf
The ratio of the cross section containing all initial state corrections  up 
to $O(a^3 L^3)$ (dotted line), 
to $O(a^3 L^2)$ (full line), 
to $O(a^3 L)$ (dash-dotted line), 
universal terms contributing to $O(a^3 L^0)$ (dashed line),
to $O(a^4 L^4)$ (long dashed line), 
normalized on all terms including 
also the $O(a^3 L)$ and $O(a^4 L^4)$ corrections to the cross section $e^+ e^- \rightarrow \gamma^*/Z^*$ in the region of the $Z$-peak
for $M_Z = 91.1876~\GeV$ \cite{PDG}.}
\label{FIG2}
\end{figure}
%-----------------------------------------------------------------------------------------------------------------
In Figures~\ref{FIG1} and \ref{FIG2} we compare the ratios of the three--loop terms with all contributions 
(\ref{EQ:EXP1}--\ref{EQ:EXP2}) in the kinematic region  of $\sqrt{s} \in [85, 95]~\GeV$.

The $O(a)$ radiative corrections are large and amount to $\sim \pm 40\%$, followed by the $O(a^2)$ corrections, still 
varying from $+15\%$ to $-7\%$, cf.~Figure~\ref{FIG1}. Already the leading term $O(a^3 {L}^3)$ yields only 
corrections at the 1\% level.
The $O(a^3)$ corrections up to the $O(a^3 L)$ term are 
significantly smaller and are illustrated in Figure~\ref{FIG2}.  
\begin{table}[H]
    \centering
    \begin{tabular}{|l|r|r|r|r|}
    \hline
    \multicolumn{1}{|c|}{} &
    \multicolumn{2}{c|}{Fixed width} &
    \multicolumn{2}{c|}{$s$ dep. width} \\
    \hline
    \multicolumn{1}{|c|}{} &   
    \multicolumn{1}{c|}{Peak} &  
    \multicolumn{1}{c|}{Width}    &
    \multicolumn{1}{c|}{Peak} & 
    \multicolumn{1}{c|}{Width} \\  
    \multicolumn{1}{|c|}{} &   
    \multicolumn{1}{c|}{(MeV)} & 
    \multicolumn{1}{c|}{(MeV)}    &
    \multicolumn{1}{c|}{(MeV)} &
    \multicolumn{1}{c|}{(MeV)} \\
    \hline
    %lowest order                            &       &     &           &          \\
    $O(\alpha)$   correction                 &   185.638   &   539.408  &   181.098  &   524.978 \\
    $O(\alpha^2 L^2)$:                       & -- 96.894   & --177.147  & -- 95.342  & --176.235 \\
    $O(\alpha^2 L)$:                         &     6.982   &    22.695  &     6.841  &    21.896 \\
    $O(\alpha^2 )$:                          &     0.176   & --  2.218  &     0.174  & --  2.001 \\
    $O(\alpha^3 L^3)$:                       &     23.265  &    38.560  &    22.968  &    38.081 \\
    $O(\alpha^3 L^2)$:                       & --   1.507  & --  1.888  & --  1.491  & --  1.881 \\
    $O(\alpha^3 L)$:                         & --   0.152  &     0.105  & --  0.151  & --  0.084 \\
    $O(\alpha^4 L^4)$:                       & --   1.857  &     0.206  & --  1.858  &     0.146 \\
    $O(\alpha^4 L^3)$:                       &      0.131  & --  0.071  &     0.132  & --  0.065 \\
    $O(\alpha^4 L^2)$:                       &      0.048  & --  0.001  &     0.048  &     0.001 \\
    $O(\alpha^5 L^5)$:                       &      0.142  & --  0.218  &     0.144  & --  0.212 \\
    $O(\alpha^5 L^4)$:                       & --   0.000  &     0.020  & --  0.001  &     0.020 \\
    $O(\alpha^5 L^3)$:                       & --   0.008  &     0.009  & --  0.008  &     0.008 \\
    $O(\alpha^6 L^6)$:                       & --   0.007  &     0.027  & --  0.007  &     0.027 \\
    $O(\alpha^6 L^5)$:                       & --   0.001  &     0.000  & --  0.001  &     0.000 \\
    
    \hline
    \end{tabular}
    \caption[]{\sf Shifts in the $Z$-mass and the width due to the different contributions to the ISR QED
    radiative corrections for a fixed width of $\Gamma_Z =  2.4952~\GeV$  and $s$-dependent width using
    $M_Z = 91.1876~\GeV$
    \cite{PDG} and $s_0 = 4 m_\tau^2$, cf.~\cite{ALEPH:2005ab}.\footnotemark}
    \label{TAB1}
    \end{table}
%%%%%%%%%%%%%%%%%%%%%%%%%%%%%%%%%%%%%%%%%%%%%%%%%%%%%%%%%%%%%%%%%%%%%%%%
%%%%%%%%%%%%%%%%%%%%%%%%%%%%%%%%%%%%%%%%%%%%%%%%%%%%%%%%%%%%%%%%%%%%%%%%
\footnotetext{In Ref.~\cite{Blumlein:2019pqb} slightly different numbers were reported for the shifts up to 
the $O(a^2)$ corrections due to the use of $a(M_Z^2)$. Here we refer to $a(m_e^2)$.}

Finally, we summarize the shifts of the $Z$ peak and the corrections to the $Z$ width, $\Gamma_Z$, by the different
orders of the ISR radiative corrections in Table~\ref{TAB1}. Here we compare the results for the fixed and the 
$s$-dependent width \cite{SDEP}. In Figures~\ref{FIG1} and \ref{FIG2} we depicted only the corrections up to 
$O(a^3)$.
The other corrections up to $O(a^6 L^5)$ are only illustrated w.r.t. its shift of the $Z$ mass and change of the 
$Z$ width since their behaviour is rather flat, yet they have an impact given the projected experimental resolutions.  
The difference to the results in Ref.~\cite{Berends:1987ab} amounts to 4 MeV in 
$\Gamma_Z$,~~\cite{Blumlein:2019pqb,QED2019}.

The relative shifts in adding the respective order can be positive or negative. at leading order $O(\alpha^k L^k)$
the level of $100~\keV$ \cite{DENT}\footnote{Statistical
accuracies of $\Delta M_Z = 0.005~\MeV~{\rm [stat]}, \Delta \Gamma_Z = 0.008~\MeV~{\rm [stat]}$ are quoted in 
\cite{DENT}.} is only undershoot at $O(\alpha^5 L^4)$. Even the 
$O(\alpha^4 L^2)$ corrections 
are of the order of $50~\keV$, while at $O(\alpha^5 L^3)$ the level of $10~\keV$ is reached.
A missing link is still the $O(\alpha^3)$ term, which can be estimated to be roughly of $O(30~\keV)$, setting the 
frame of accuracy which is currently reached for the initial state corrections.
The QED corrections to 3--loop order are still somewhat larger than the experimental resolution at the the 
FCC\_ee \cite{FCCEE} making the inclusion of also higher order subleading corrections necessary.

Furthermore, we remark that we have calculated the inclusive ISR corrections only, assuming that the 
experimental data are extrapolated to the full phase space and only a cut in $s'$ is considered.
The experimental requirements may be more ambitious, requiring more differential radiative corrections in the 
future. Due to different cuts, the corrections will turn out to be different and one has to carefully study 
all the cut dependencies. For a recent summary see \cite{Jadach:2019huc}.
From the size of the corrections it seems that 3- to 4-loop corrections have there to 
be provided too. 

Numerical implementations for harmonic polylogarithms in {\tt Fortran} needed for the radiators in $z$ space are 
given in 
\cite{Gehrmann:2001pz} and \cite{Ablinger:2018sat}, respectively. 
%--------------------------------------------------------------------------------------------------------
\section{Conclusions}
\label{sec:6}
%--------------------------------------------------------------------------------------------------------

\vspace*{1mm}
\noindent
We have calculated the QED initial state corrections to the annihilation process $e^+e^- \rightarrow 
\gamma^*/Z^*$ up to the terms $O(\alpha^6 L^5)$. They come next to the recently completed $O(\alpha^2)$ 
corrections 
\cite{Blumlein:2019srk,Blumlein:2019pqb,QED2019} and the well--known universal corrections $O((\alpha L)^k)$, 
which were
known to fifth order in explicit form in the non--singlet and singlet case \cite{Skrzypek:1992vk,Jezabek:1992bx,
Przybycien:1992qe,Blumlein:1996yz,Arbuzov:1999cq,Arbuzov:1999uq,Blumlein:2004bs,Blumlein:2007kx}. 
Here we included the first three logarithmic terms for the orders $O(\alpha^3)$ to $O(\alpha^5)$. The radiators
are given by convolutions of splitting functions, the contributions to the Wilson coefficient of the massless 
Drell--Yan process and massive operator matrix elements. For the present corrections the massive OME 
$\Gamma_{\gamma e}^{(1)}$ had to be calculated newly. The other massive OMEs were given in \cite{Blumlein:2011mi} 
before. The corrections calculated in the present paper can be expressed in terms of harmonic 
polylogarithms, if one also allows for the argument $(1-x)$ in case of harmonic polylogarithms containing an  
index $i = -1$. In Mellin space the radiators can be represented in terms of harmonic sums and generalized 
harmonic sums. The present corrections may still miss terms of $O(30~\keV)$ for both $\delta M_Z$ and $\delta 
\Gamma_Z$,  which can be further improved by  calculating the yet missing terms.

It is needless to say that by performing Mellin convolutions, using the quantities calculated in the present 
paper and the massless quantities available in the literature, one is now in the position 
to calculate all corrections of $O(\alpha^k L^k), O(\alpha^k L^{k-1})$ and $O(\alpha^k L^{k-2})$ for $k \geq 5$
straightforwardly. For the projected experimental accuracies in inclusive measurements at the FCC\_ee 
they may not be needed beyond the orders already obtained.

\appendix
%--------------------------------------------------------------------------------------------------------
\section{The Singularities of the Radiators in \bf N space}
\label{sec:A}
%--------------------------------------------------------------------------------------------------------

\vspace*{1mm}
\noindent
For the use of complex contour integrals to calculate the radiative corrections the position of the 
singularities of the radiators in Mellin $N$ space in the complex plane have to be known.

In the case of harmonic sums, except for $S_1(N)$, it is known from their representation in terms of factorial 
series \cite{FACTS} that their singularities are given by the set $\{-n~|~n \in \mathbb{N}\}$. The sum $S_1(N)$
has a representation by the di-gamma function $\psi(N)$ for which the same holds. 

A factorial series is given by
%-------------------------------------------------------------------------------------
\begin{eqnarray}
\label{eq:FS}
\Omega(x) = a_0 + \sum_{k=0}^\infty \frac{a_{k+1} k!}{x(x+1) ... (x+n)},~~\text{with}~~a_i \in \mathbb{C}.
\end{eqnarray}
%-------------------------------------------------------------------------------------
The question to be answered is, which functions can be expanded into factorial series. The structure of 
(\ref{eq:FS}) then provides the singularity structure for $x \in \mathbb{C}$. If a function $F(x)$ has the
representation 
%-------------------------------------------------------------------------------------
\begin{eqnarray}
\label{eq:FS1}
F(x) = \int_0^1~dt~t^x~\varphi(t)
\end{eqnarray}
%------------------------------------------------------------------------------------- 
and the function $\varphi(t)$  can be expanded into a Taylor series in $(1-t)$, partial integration
will then lead to a factorial series.

The radiators are expressed in terms of the following monomials
%-------------------------------------------------------------------------------------
\begin{eqnarray}
\label{eq:FS2}
\frac{\HA_{\vec{a}}(\xi)}{z^k (1-z)^l (1+z)^m},~~~l,m \in \mathbb{N},~~~k \in \{0,1\} 
\end{eqnarray}
%------------------------------------------------------------------------------------- 
and $\xi \in \{z, 1-z\}$, with $a_i \in \{0,1,-1\}$. To perform the Mellin transform of (\ref{eq:FS2})
suitable regularizations have to be chosen. Since the Mellin transform of the complete radiator
exists and is unique, these partly arbitrary regularization terms add up to zero.

Now one has to assure that $\HA_{\vec{a}}(\xi)$ can be expanded into a Taylor series in the variable $(1-z)$.
This will require in case to change the argument from 
%-------------------------------------------------------------------------------------
\begin{eqnarray}
\label{eq:FS3a}
z \leftrightarrow (1-z),
\end{eqnarray}
%------------------------------------------------------------------------------------- 
which is a valid operation on $\HA_{\vec{a}}(\xi)$ on the expense of introducing the letter 
%-------------------------------------------------------------------------------------
\begin{eqnarray}
\label{eq:FS3}
\frac{1}{2-z}.
\end{eqnarray}
%------------------------------------------------------------------------------------- 
The structure in (\ref{eq:FS2}) can be generated by applying the following differential operator
%-------------------------------------------------------------------------------------
\begin{eqnarray}
\label{eq:FS4}
\frac{\HA_{\vec{a}}(z)}{z^k (1-z)^l (1+z)^m} = \frac{1}{z^k} \frac{d^{m+l}}{dx^{m+l}} \HA_{\vec{c}_1, 
\vec{c}_2,\vec{a}}(z),
\end{eqnarray}
%-------------------------------------------------------------------------------------
with 
%-------------------------------------------------------------------------------------
\begin{eqnarray}
\label{eq:FS5}
\{c_{11}, ..., c_{1m}\} = \{1,1, ... ,1\},~~~~~\{c_{21}, ..., c_{2l}\} = \{-1,-1, ... ,-1\}.
\end{eqnarray}
%-------------------------------------------------------------------------------------
Furthermore, one has
%-------------------------------------------------------------------------------------
\begin{eqnarray}
\label{eq:FS6}
\int_0^1 dz~z^N~\frac{d}{dz} \HA_{\vec{b}}(z) = \HA_{\vec{b}}(1) -
N \int_0^1 dz~z^{N-1}~\HA_{\vec{b}}(z).
\end{eqnarray}
%-------------------------------------------------------------------------------------
In this way and by the argument mapping (\ref{eq:FS3a}) one arrives at valid functions $\varphi(t)$ allowing to 
expand into a factorial series. The above construction has now to be applied to all radiators and one finds
the set of singularities to be a subset of $\{-n~|~n \in \mathbb{N} \cup \{-1\}\}$.

%--------------------------------------------------------------------------------------------------------
\section{The Radiators in N space}
\label{sec:B}
%--------------------------------------------------------------------------------------------------------

\vspace*{1mm}
\noindent
In the following we list all radiators which were calculated in the present paper in Mellin $N$ space for the 
use in Mellin space programs. The analytic continuation of the respective harmonic sums can be performed as 
described in Refs.~\cite{ANCONT,Blumlein:2009ta}. The package {\tt HarmonicSums} allows to derive the asymptotic
representation of these coefficients. Their recursion relations follow from the ones of the harmonic and 
generalized 
harmonic sums. As has been shown, there are no singularities right to $N=1$, which allows to perform the contour 
integral to $z$ space with the usual contour in the singlet--case, see e.g.~\cite{Blumlein:1997em}, surrounding the
singularities of the expression in the complex plane, cf.~Appendix~\ref{sec:A}.

The radiators $R_{ij}$ are related to the expansion coefficients from $c_{3,3}$ to $c_{6,5}$, also labeling the 
difference term (\ref{eq:DIFF}) 
between the axial--vector and vector contributions, by
%-------------------------------------------------------------------------------------
\begin{eqnarray}
R_{ij}(N)  = \Mvec[c_{ij}(x)](N).
\end{eqnarray}
%-------------------------------------------------------------------------------------
Alternating sums can be rewritten in terms of Mellin transforms such that the contribution
due to $\ln(2)$ terms cancel, cf. \cite{Blumlein:2006mh}. We have also checked that the evanescent poles present 
in the above 
radiators  up to $1/(N-4)$, cancel, leaving the rightmost pole $1/(N-1)$. 

The analytic continuation is performed from the even integers. It can be obtained in the analyticity region for $N 
\rightarrow \mathbb{C}$ for both the harmonic 
sums and generalized harmonic sums expressing them for large values of $|N|$ by their asymptotic 
expansion and by using the recursion relations to map to finite values of $N \in \mathbb{C}$, \cite{Blumlein:2009ta}.

The radiators in $N$ space are given by
%-------------------------------------------------------------------------------------
\input{nspace2.tex}

%-------------------------------------------------------------------------------------

\noindent
The polynomials $Q_k|_{k=1}^{213}$ are to long to be displayed here and they are given in an ancillary file to 
this paper.

\vspace*{5mm}
\noindent 
{\bf Acknowledgments.}\\
We would like to thank P.~Marquard and C.~Schneider for discussions. This project has received funding from 
the European Union's Horizon 2020 research and innovation programme under the Marie Sk\/{l}odowska-Curie grant 
agreement No. 764850, SAGEX, COST action CA16201: Unraveling new physics at the LHC through the precision 
frontier, and Austrian Science Fund (FWF) grant SFB F50 (F5009-N15). The large formulae were tyosetted by 
using {\tt SigmaToTeX} of Ref.~\cite{SIG1,SIG2}. The diagrams have been drawn using {\tt Axodraw} 
\cite{Vermaseren:1994je}.
%-------------------------------------------------------------------------------------

%-----------------------------------------------------------------------------------
\end{document}

%% file: formu2.tex
%--------------------------------------------------------------------------------------------------------

%--------------------------------------------------------------------------------------------------------

%% file: xspaceD.tex
%-----------------------------------------------------------------------------------------------------------
\begin{eqnarray}
c_{3,3}^{\delta} &=&
\Biggl[
  \frac{572}{9}
        -\frac{704 \zeta_2}{3}
        +\frac{512 \zeta_3}{3}
\Biggr] \delta(1-z)
        \\
%-------
c_{3,2}^{\delta} &=&
\Biggl[
        -\frac{2774}{9}
        +\frac{7424 \zeta_2}{9}
        -256 \zeta_2^2
        +\frac{32 \zeta_3}{3}
\Biggr] \delta(1-z)
\\
%-------
c_{3,1}^{\delta} &=& 
\Biggl[
        \frac{28889}{27}
        +\Biggl(
                -\frac{23968}{27}
                -576 \ln(2)
                +320 \zeta_3
        \Biggr) \zeta_2
        +\frac{5536}{15} \zeta_2^2
        -808 \zeta_3
        -480 \zeta_5
\Biggr] \delta(1-z)
\\
%-------
c_{4,4}^{\delta} &=&
\Biggl[
        \frac{1430}{9}
        -\frac{27328 \zeta_2}{27}
        +\frac{512 \zeta_2^2}{5}
        +\frac{4096 \zeta_3}{3}
\Biggr] \delta(1-z)
\\
%-------
c_{4,3}^{\delta} &=& 
\Biggl[
        -
        \frac{27938}{27}
        +\Biggl(
                \frac{134032}{27}
                -\frac{512 \zeta_3}{3}
        \Biggr) \zeta_2
        -\frac{24832}{15} \zeta_2^2
        -\frac{30304}{9} \zeta_3
\Biggr] \delta(1-z)
\\
%-------
c_{4,2}^{\delta} &=& 
\Biggl[
        \frac{707717}{162}
        +\Biggl(
                -\frac{908020}{81}
                -2112 \ln(2)
                -\frac{6016 \zeta_3}{3}
        \Biggr) \zeta_2
        +\frac{32}{15} (1883+1440 \ln(2)) \zeta_2^2
\nonumber\\ &&        
+\frac{3072}{5} \zeta_2^3
        +672 \zeta_3
        +1152 \zeta_3^2
        +\frac{4864}{3} \zeta_5
\Biggr] \delta(1-z).
\end{eqnarray}
%-----------------------------------------------------------------------------------------------------------

%% file: xspaceP.tex
%--------------------------------------------------------------------------------------------------------
\begin{eqnarray}
c_{3,3}^{+} &=&
        \Biggl(
                \frac{5744}{27}
                -256 \zeta_2
        \Biggr) {\cal D}_0
        +\frac{1408}{3} {\cal D}_1
        +256 {\cal D}_2
\\
%------
c_{3,2}^{+} &=&
        \Biggl(
                -
                \frac{23608}{27}
                +832 \zeta_2
                +384 \zeta_3
        \Biggr) {\cal D}_0
        +\Biggl(
                -\frac{4480}{3}
                +512 \zeta_2
        \Biggr) {\cal D}_1
        -512 {\cal D}_2
\\
%----------
c_{3,1}^{+} &=& 
        \Biggl(
                \frac{164608}{81}
                +\Biggl(
                        -\frac{7696}{9}-768 \ln(2) \Biggr) \zeta_2
                -\frac{768 \zeta_2^2}{5}
                -320 \zeta_3
        \Biggr) {\cal D}_0
        +\Biggl(
                \frac{48832}{27}
                -1536 \zeta_2
        \Biggr){\cal D}_1
\nonumber\\ &&      
  +\frac{2432}{9} {\cal D}_2
        +\frac{2048}{9} {\cal D}_3
\\
%----------
c_{4,4}^{+} &=&
        \Biggl(
                \frac{17792}{27}
                -2048 \zeta_2
                +\frac{4096 \zeta_3}{3}
        \Biggr) {\cal D}_0
        +\Biggl(
                \frac{54656}{27}
                -2048 \zeta_2
        \Biggr) {\cal D}_1
        +2048 {\cal D}_2
        +\frac{2048}{3} {\cal D}_3
\\
%----------
c_{4,3}^{+} &=&
        \Biggl(
                -\frac{308320}{81}
                +\frac{221056 \zeta_2}{27}
                -2048 \zeta_2^2
                +\frac{1024 \zeta_3}{3}
        \Biggr) {\cal D}_0
        +\Biggl(
                -\frac{263488}{27}
                +\frac{20992 \zeta_2}{3}
                +3072 \zeta_3
        \Biggr) {\cal D}_1
\nonumber\\ &&      
  +\Biggl(
                -\frac{22016}{3}
                +2048 \zeta_2
        \Biggr) {\cal D}_2
        -\frac{4096 {\cal D}_3}{3}
\\
%----------
c_{4,2}^{+} &=& 
          \Biggl(
                \frac{3135184}{243}
                +\Biggl(
                        -\frac{333728}{27}
                        -5120 \ln(2)
                        +2560 \zeta_3
                \Biggr) \zeta_2
                +\frac{19712}{5} \zeta_2^2
                -\frac{72128}{9} \zeta_3
                -3840 \zeta_5
        \Biggr) {\cal D}_0 
\nonumber\\ &&
+        
\Biggl(
                \frac{1979456}{81}
                +\Biggl(
                        -\frac{36224}{3}-6144 \ln(2) \Biggr) \zeta_2
                -\frac{6144 \zeta_2^2}{5}
                +\frac{5632 \zeta_3}{3}
        \Biggr) {\cal D}_1
\nonumber\\ &&        
+\Biggl(
                \frac{99712}{9}
                -\frac{32768 \zeta_2}{3}
        \Biggr) {\cal D}_2
        +\frac{38912}{27} {\cal D}_3
        +\frac{10240}{9} {\cal D}_4,
%----------
\end{eqnarray}
%--------------------------------------------------------------------------------------------------------

%% file: xspace1.tex
%------------------------------------------------------------------------------------------------------------------------

%--------------------------------------------------------------------------------------------------------

%% file: nspace2.tex
%---------------------------------------------------------------------------------------------------------

%---------------------------------------------------------------------------------------------------------

%% file: paper1.bbl
\begin{thebibliography}{99}
%-------------------------------------------------------------------------------------
%
%[1]
\bibitem{ILC}
%\bibitem{Accomando:1997wt}
  E.~Accomando {\it et al.} %[ECFA/DESY LC Physics Working Group],
  %``Physics with $e^{+} e^{-}$ linear colliders,''
  Phys.\ Rept.\  {\bf 299} (1998) 1--78
%  doi:10.1016/S0370-1573(97)00086-0
  [hep-ph/9705442];\\
%%CITATION = doi:10.1016/S0370-1573(97)00086-0;%% 
%\bibitem{AguilarSaavedra:2001rg}
  J.A.~Aguilar-Saavedra {\it et al.} %[ECFA/DESY LC Physics Working Group],
%  {\it TESLA: The Superconducting electron positron linear collider with an integrated x-ray laser laboratory. 
%  Technical design report. Part 3. Physics at an e+ e- linear collider,''
  hep-ph/0106315;\\
  %%CITATION = HEP-PH/0106315;%%
{\sf International Linear Collider Reference Design Report}, 
ILC-REPORT-2007-001, Eds. J. Brau, Y. Okada, and N. Walker; Vol.~1--4.
\\
%\bibitem{Djouadi:2007ik}
  G.~Aarons {\it et al.} [ ILC Collaboration ],
  {\sf International Linear Collider Reference Design Report Volume 2: Physics At The ILC},
  [arXiv:0709.1893 [hep-ph]];\\
%%CITATION = ARXIV:0709.1893;%%
{\tt http://www.linearcollider.org/ILC}
%-------------------------------------------------------------------------------------
%
%[2]
\bibitem{Aihara:2019gcq}
  H.~Aihara {\it et al.} [Linear Collider Collaboration],
  {\it The International Linear Collider. A Global Project},
  arXiv:1901.09829 [hep-ex].
  %%CITATION = ARXIV:1901.09829;%%
%-------------------------------------------------------------------------------------
%
%[3]
\bibitem{Mnich:2019}
  J.~Mnich,
  {\it The International Linear Collider: Prospects and Possible Timelines},
  arXiv: 1901.10206 [hep-ex].
  %%CITATION = ARXIV:1901.10206;%%
%-------------------------------------------------------------------------------------
%
%[4]
\bibitem{CLIC}
%\bibitem{vanderMeer:1988yu}
  S.~van der Meer,
  {\sf The CLIC Project and the Design for an $e^+ e^-$ Collider}, CLIC-NOTE-68,
  (1988);\\
R.W. Assmann et al., CLIC Study Team, {\sf A 3 TeV $e^+ e^-$ Linear Collider
Based on CLIC Technology}, CERN 2000-008;\\
%\bibitem{Accomando:2004sz}  
  E.~Accomando {\it et al.}  [CLIC Physics Working Group],
  {\it Physics at the CLIC multi-TeV linear collider,}
  arXiv:hep-ph/0412251;\\
  %%CITATION = HEP-PH/0412251;%%
%\bibitem{Roloff:2018dqu}
  P.~Roloff {\it et al.} [CLIC and CLICdp Collaborations],
  {\it The Compact Linear e$^+$e$^-$ Collider (CLIC): Physics Potential},   
  arXiv:1812.07986 [hep-ex]. 
  %%CITATION = ARXIV:1812.07986;%%
%-------------------------------------------------------------------------------------
%
%[5]
\bibitem{FCCEE}
{\tt http://tlep.web.cern.ch/}
%-------------------------------------------------------------------------------------
%
%[6]
\bibitem{Delahaye:2019omf}
  J.P.~Delahaye {\it et al.},
  {\it Muon Colliders},
  arXiv:1901.06150 [physics.acc-ph].
%-------------------------------------------------------------------------------------
%
%[7]
\bibitem{ALEPH:2005ab}
  S.~Schael {\it et al.} 
  %[ALEPH and DELPHI and L3 and OPAL and SLD Collaborations and LEP Electroweak Working Group and SLD Electroweak Group and SLD Heavy Flavour Group],
  %``Precision electroweak measurements on the $Z$ resonance,''
  Phys.\ Rept.\  {\bf 427} (2006) 257--454
%  doi:10.1016/j.physrep.2005.12.006
  [hep-ex/0509008].
  %%CITATION = doi:10.1016/j.physrep.2005.12.006;%%
%-------------------------------------------------------------------------------------
%
%[8]
\bibitem{Blumlein:2019srk}
  J.~Bl\"umlein, A.~De Freitas, C.G.~Raab and K.~Sch\"onwald,
  %``The $O(\alpha^2)$ Initial State QED Corrections to $e^+e^-$ Annihilation to a Neutral Vector Boson Revisited,''
  Phys.\ Lett.\ B {\bf 791} (2019) 206--209
%  doi:10.1016/j.physletb.2019.02.038
  [arXiv:1901.08018 [hep-ph]].
  %%CITATION = doi:10.1016/j.physletb.2019.02.038;%%
%-------------------------------------------------------------------------------------
%
%[9]
\bibitem{Blumlein:2019pqb}
  J.~Bl\"umlein, A.~De Freitas, C.G.~Raab and K.~Sch\"onwald,
  %``The effects of $O(\alpha^2)$ initial state QED corrections to $e^+e^- \rightarrow \gamma^*/Z^*$ at very high luminosity colliders,''
  Phys.\ Lett.\ B {\bf 801} (2020) 135196
%  doi:10.1016/j.physletb.2019.135196
  [arXiv:1910.05759 [hep-ph]].
  %%CITATION = doi:10.1016/j.physletb.2019.135196;%%
%-------------------------------------------------------------------------------------
%
%[10]
\bibitem{QED2019}
J.~Bl\"umlein, A.~De Freitas, C.G.~Raab and K.~Sch\"onwald, {\it The $O(\alpha^2)$ Initial State QED Corrections to 
$e^+e^- \rightarrow \gamma^*/{Z^{0}}^*$}, DESY 18--196.
%-------------------------------------------------------------------------------------
%
%[11]
\bibitem{Berends:1987ab}
  F.A.~Berends, W.L.~van Neerven and G.J.H.~Burgers,
  %``Higher Order Radiative Corrections at LEP Energies,''
  Nucl.\ Phys.\ B {\bf 297} (1988) 429--478. Erratum: [Nucl.\ Phys.\ B {\bf 304} (1988) 921--922].
%  doi:10.1016/0550-3213(88)90313-6, 10.1016/0550-3213(88)90662-1
  %%CITATION = doi:10.1016/0550-3213(88)90313-6, 10.1016/0550-3213(88)90662-1;%%
%-------------------------------------------------------------------------------------
%
%[12]
\bibitem{Blumlein:2011mi}
  J.~Bl\"umlein, A.~De Freitas and W.L.~van Neerven,
  %``Two-loop QED Operator Matrix Elements with Massive External Fermion Lines,''
  Nucl.\ Phys.\ B {\bf 855} (2012) 508--569
%  doi:10.1016/j.nuclphysb.2011.10.009
  [arXiv:1107.4638 [hep-ph]].
  %%CITATION = doi:10.1016/j.nuclphysb.2011.10.009;%%
%-------------------------------------------------------------------------------------
%
%[13]
\bibitem{Skrzypek:1992vk}
M.~Skrzypek,
%``Leading logarithmic calculations of QED corrections at LEP,''
Acta Phys.\ Polon.\  {\bf B23} (1992) 135--172.
%%CITATION = APPOA,B23,135;%%
%----------------------------------------------------------------------------------------------------------------------------------
%
%[14]
\bibitem{Jezabek:1992bx}
M.~Jezabek,
%``A Perturbative solution to Gribov-Lipatov equation,''
Z.\ Phys.\  {\bf C56} (1992) 285--288.
%%CITATION = ZEPYA,C56,285;%% 
%----------------------------------------------------------------------------------------------------------------------------------
%
%[15]
\bibitem{Przybycien:1992qe}
  M.~Przybycien,
  %``A Fifth order perturbative solution to the Gribov-Lipatov equation,''
  Acta Phys.\ Polon.\ B {\bf 24} (1993) 1105--1114
  [hep-th/9511029].
  %%CITATION = HEP-TH/9511029;%%
%----------------------------------------------------------------------------------------------------------------------------------
%
%[16]
\bibitem{Blumlein:1996yz}
  J.~Bl\"umlein, S.~Riemersma and A.~Vogt,
  %``On the resummation of the alpha ln**2 z terms for QED corrections to deep inelastic e p scattering and e+ e- annihilation,''
  Eur.\ Phys.\ J.\ C {\bf 1} (1998) 255--259
%  doi:10.1007/BF01245815
  [hep-ph/9611214].
  %%CITATION = doi:10.1007/BF01245815;%%
%-------------------------------------------------------------------------------------
%
%[17]
\bibitem{Arbuzov:1999cq}
  A.B.~Arbuzov,
  %``Nonsinglet splitting functions in QED,''
  Phys.\ Lett.\ B {\bf 470} (1999) 252--258
%  doi:10.1016/S0370-2693(99)01290-3
  [hep-ph/9908361].
  %%CITATION = doi:10.1016/S0370-2693(99)01290-3;%%
%----------------------------------------------------------------------------------------------------------------------------------
%
%[18]
\bibitem{Arbuzov:1999uq}
  A.B.~Arbuzov,
  %``Higher order pair corrections to electron positron annihilation,''
  JHEP {\bf 0107} (2001) 043
%  doi:10.1088/1126-6708/2001/07/043
  [hep-ph/9907500].
  %%CITATION = doi:10.1088/1126-6708/2001/07/043;%%
%----------------------------------------------------------------------------------------------------------------------------------
%
%[19]
\bibitem{Blumlein:2004bs}
  J.~Bl\"umlein and H.~Kawamura,
  %``Universal higher order QED corrections to polarized lepton scattering,''
  Nucl.\ Phys.\ B {\bf 708} (2005) 467--510
%  doi:10.1016/j.nuclphysb.2004.12.001
  [hep-ph/0409289].
  %%CITATION = doi:10.1016/j.nuclphysb.2004.12.001;%%
%-------------------------------------------------------------------------------------
%
%[20]
\bibitem{Blumlein:2007kx}
  J.~Bl\"umlein and H.~Kawamura,
  %``Universal higher order singlet QED corrections to unpolarized lepton
  %scattering,''
  Eur.\ Phys.\ J.\  C {\bf 51} (2007) 317--333
  [arXiv:hep-ph/0701019].
  %%CITATION = EPHJA,C51,317;%%
%----------------------------------------------------------------------------------------------------------------------------------
%
%[21]
\bibitem{Blumlein:2002fy}
  J.~Bl\"umlein and H.~Kawamura,
  %``O(alpha**2 L) radiative corrections to deep inelastic ep scattering,''
  Phys.\ Lett.\ B {\bf 553} (2003) 242--250
%  doi:10.1016/S0370-2693(02)03194-5
  [hep-ph/0211191];\\
  %%CITATION = doi:10.1016/S0370-2693(02)03194-5;%%
%\bibitem{Arbuzov:2019hcg}
  A.B.~Arbuzov,
  %``Leading and Next-to-Leading Logarithmic Approximations in Quantum Electrodynamics,''
  Phys.\ Part.\ Nucl.\  {\bf 50} (2019) no.6,  721--825.
%  doi:10.1134/S1063779619060029
%%CITATION = doi:10.1134/S1063779619060029;%%
%-----------------------------------------------------------------------------------
%
%[22]
\bibitem{NLOand}
%\bibitem{Floratos:1977au}
  E.G.~Floratos, D.A.~Ross and C.T.~Sachrajda,
  %``Higher Order Effects In Asymptotically Free Gauge Theories: The Anomalous
  %Dimensions Of Wilson Operators,''
  Nucl.\ Phys.\  B {\bf 129} (1977) 66--88
  [Erratum-ibid.\  B {\bf 139} (1978) 545--546];
  %%CITATION = NUPHA,B129,66;%%
%\bibitem{Floratos:1978ny}
%  E.G.~Floratos, D.A.~Ross and C.T.~Sachrajda,
  %``Higher Order Effects In Asymptotically Free Gauge Theories. 2. Flavor
  %Singlet Wilson Operators And Coefficient Functions,''
  Nucl.\ Phys.\  B {\bf 152} (1979) 493--520;\\
  %%CITATION = NUPHA,B152,493;%%;
%\bibitem{GonzalezArroyo:1979df}
  A.~Gonzalez-Arroyo, C.~Lopez and F.~J.~Yndurain,
  %``Second Order Contributions to the Structure Functions in Deep Inelastic Scattering. 1. Theoretical Calculations,''
  Nucl.\ Phys.\ B {\bf 153} (1979) 161--186;\\
%  doi:10.1016/0550-3213(79)90596-0, 10.1016/0550-3213(79)90466-8
  %%CITATION = doi:10.1016/0550-3213(79)90596-0, 10.1016/0550-3213(79)90466-8;%%
%\bibitem{GonzalezArroyo:1979he}
  A.~Gonzalez-Arroyo and C.~Lopez,
  %``Second Order Contributions To The Structure Functions In Deep Inelastic
  %Scattering. 3. The Singlet Case,''
  Nucl.\ Phys.\  B {\bf 166} (1980) 429--459;\\
  %%CITATION = NUPHA,B166,429;%%
%\bibitem{Floratos:1981hs}
  E.G.~Floratos, C.~Kounnas and R.~Lacaze,
  %``Higher Order QCD Effects In Inclusive Annihilation And Deep Inelastic
  %Scattering,''
  Nucl.\ Phys.\  B {\bf 192} (1981) 417--462;\\
  %%CITATION = NUPHA,B192,417;%%
%\bibitem{Curci:1980uw}
  G.~Curci, W.~Furmanski and R.~Petronzio,
  %``Evolution Of Parton Densities Beyond Leading Order: The Nonsinglet Case,''
  Nucl.\ Phys.\  B {\bf 175} (1980) 27--92;\\
  %%CITATION = NUPHA,B175,27;%%
%\bibitem{Furmanski:1980cm}
  W.~Furmanski and R.~Petronzio,
  %``Singlet Parton Densities Beyond Leading Order,''
  Phys.\ Lett.\  B {\bf 97} (1980) 437--442;\\
  %%CITATION = PHLTA,B97,437;%%
%\bibitem{Hamberg:1991qt}
  R.~Hamberg and W.L.~van Neerven,
  %``The Correct renormalization of the gluon operator in a covariant gauge,''
  Nucl.\ Phys.\  B {\bf 379} (1992) 143--171;\\
  %%CITATION = NUPHA,B379,143;%%
%\bibitem{Mertig:1995ny}
  R.~Mertig and W.L.~van Neerven,
  %``The Calculation of the two loop spin splitting functions P(ij)(1)(x),''
  Z.\ Phys.\ C {\bf 70} (1996) 637--654
%  doi:10.1007/s002880050138
  [hep-ph/9506451];\\
  %%CITATION = doi:10.1007/s002880050138;%%
%\bibitem{Vogelsang:1995vh}
  W.~Vogelsang,
  %``A Rederivation of the Spin-dependent Next-to-leading Order Splitting
  %Functions,''
  Phys.\ Rev.\  D {\bf 54} (1996) 2023--2029
  [hep-ph/9512218];
  %%CITATION = PHRVA,D54,2023;%%
%\bibitem{Vogelsang:1996im}
%  W.~Vogelsang,
  %``The spin-dependent two-loop splitting functions,''
  Nucl.\ Phys.\  B {\bf 475} (1996) 47--72
  [hep-ph/9603366];\\
  %%CITATION = NUPHA,B475,47;%%
%\bibitem{Ellis:1996nn}
  R.K.~Ellis and W.~Vogelsang,
  %``The evolution of parton distributions beyond leading order: the singlet
  %case,''
  arXiv:hep-ph/9602356;\\
  %%CITATION = HEP-PH/9602356;%%
%\bibitem{Moch:1999eb}
  S.~Moch, J.A.M.~Vermaseren,
  %``Deep inelastic structure functions at two loops,''
  Nucl.\ Phys.\  {\bf B573 } (2000)  853--907
  [hep-ph/9912355];\\
%\bibitem{Moch:2004pa}
  S.~Moch, J.~A.~M.~Vermaseren and A.~Vogt,
  %``The Three loop splitting functions in QCD: The Nonsinglet case,''
  Nucl.\ Phys.\ B {\bf 688} (2004) 101--134
%  doi:10.1016/j.nuclphysb.2004.03.030
  [hep-ph/0403192];\\
  %%CITATION = doi:10.1016/j.nuclphysb.2004.03.030;%%
%\bibitem{Vogt:2004mw}
  A.~Vogt, S.~Moch and J.~A.~M.~Vermaseren,
  %``The Three-loop splitting functions in QCD: The Singlet case,''
  Nucl.\ Phys.\ B {\bf 691} (2004) 129--181
%  doi:10.1016/j.nuclphysb.2004.04.024
  [hep-ph/0404111];\\
  %%CITATION = doi:10.1016/j.nuclphysb.2004.04.024;%%
%\bibitem{Moch:2014sna}
  S.~Moch, J.A.M.~Vermaseren and A.~Vogt,
  %``The Three-Loop Splitting Functions in QCD: The Helicity-Dependent Case,''
  Nucl.\ Phys.\ B {\bf 889} (2014) 351--400
%  doi:10.1016/j.nuclphysb.2014.10.016
  [arXiv:1409.5131 [hep-ph]];\\
  %%CITATION = doi:10.1016/j.nuclphysb.2014.10.016;%%
%\bibitem{Ablinger:2017tan}
  J.~Ablinger, A.~Behring, J.~Bl\"umlein, A.~De Freitas, A.~von Manteuffel and C.~Schneider,
  %``The three-loop splitting functions $P_{qg}^{(2)}$ and $P_{gg}^{(2, N_F)}$,''
  Nucl.\ Phys.\ B {\bf 922} (2017) 1--40
%  doi:10.1016/j.nuclphysb.2017.06.004
  [arXiv:1705.01508 [hep-ph]];\\
  %%CITATION = doi:10.1016/j.nuclphysb.2017.06.004;%%
%\bibitem{Behring:2019tus}
  A.~Behring, J.~Bl\"umlein, A.~De Freitas, A.~Goedicke, S.~Klein, A.~von Manteuffel, C.~Schneider and K.~Sch\"onwald,
  %``The Polarized Three-Loop Anomalous Dimensions from On-Shell Massive Operator Matrix Elements,''
  Nucl.\ Phys.\ B {\bf 948} (2019) 114753
%  doi:10.1016/j.nuclphysb.2019.114753
  [arXiv:1908.03779 [hep-ph]].
  %%CITATION = doi:10.1016/j.nuclphysb.2019.114753;%%
%----------------------------------------------------------------------------------------------------------------------------------
%
%[23]
\bibitem{Hamberg:1990np}
  R.~Hamberg, W.L.~van Neerven and T.~Matsuura,
  %``A complete calculation of the order $\alpha-s^{2}$ correction to the Drell-Yan $K$ factor,''
  Nucl.\ Phys.\ B {\bf 359} (1991) 343--405
   Erratum: [Nucl.\ Phys.\ B {\bf 644} (2002) 403--404].
%  doi:10.1016/S0550-3213(02)00814-3, 10.1016/0550-3213(91)90064-5
  %%CITATION = doi:10.1016/S0550-3213(02)00814-3, 10.1016/0550-3213(91)90064-5;%%
%-------------------------------------------------------------------------------------
%
%[24]
\bibitem{Harlander:2002wh}
  R.V.~Harlander and W.B.~Kilgore,
  %``Next-to-next-to-leading order Higgs production at hadron colliders,''
  Phys.\ Rev.\ Lett.\  {\bf 88} (2002) 201801
%  doi:10.1103/PhysRevLett.88.201801
  [hep-ph/0201206].
  %%CITATION = doi:10.1103/PhysRevLett.88.201801;%%
%-------------------------------------------------------------------------------------
%
%[25]
\bibitem{Remiddi:1999ew}
E.~Remiddi and J.A.M.~Vermaseren,
%{\it {Harmonic polylogarithms}},
{{ Int. J. Mod. Phys.}
  {\bfseries A15} (2000) 725--754}
{{[hep-ph/9905237]}}.
%%CITATION = HEP-PH/9905237;%%.
%-------------------------------------------------------------------------------------------------------
%
%[26]
\bibitem{NIELSEN}
N. Nielsen, 
%Der Eulersche Dilogarithmus und seine Verallgemeinerungen, 
Nova Acta
Leopold. {\bf XC} (1909) Nr. 3, 125--211;\\
%\bibitem{Kolbig:1983qt}
  K.S.~K\"olbig,
  %``Nielsen's Generalized Polylogarithms,''
  SIAM J.\ Math.\ Anal.\  {\bf 17} (1986) 1232--1258.
%  doi:10.1137/0517086
  %%CITATION = doi:10.1137/0517086;%%
%------------------------------------------------------------------------
%
%[27]
\bibitem{CLPOLY}
%\bibitem{Devoto:1983tc}
  A.~Devoto and D.~W.~Duke,
  %``Table Of Integrals And Formulae For Feynman Diagram Calculations,''
  Riv.\ Nuovo Cim.\  {\bf 7N6} (1984) 1--39;\\
  %%CITATION = RNCIB,7N6,1;%%
%---------
%\bibitem{LEWIN}
L. Lewin, {\sf Dilogarithms and associated functions} (Macdonald, London, 1958);\\
L. Lewin, {\sf Polylogarithms and Associated Functions}, (North Holland, Amsterdam, 1981).
%------------------------------------------------------------------------
%
%[28]
\bibitem{BDJ}
M. B\"ohm, A. Denner, and H. Joos, {\sf Gauge Theories of the Strong and
Electroweak Interaction}, (B.G. Teubner, Stuttgart, 2001);\\
%\bibitem{Denner:2019vbn}
  A.~Denner and S.~Dittmaier,
  {\it Electroweak Radiative Corrections for Collider Physics},
  arXiv:1912.06823 [hep-ph].
  %%CITATION = ARXIV:1912.06823;%%
%------------------------------------------------------------------------
%
%[29]
\bibitem{WB}
W.J.P.~Beenakker, {\sf Electroweak corrections: techniques and applications}, PhD Thesis, (Leiden University, 1989).
%------------------------------------------------------------------------
%
%[30]
\bibitem{DISTR}
K. Yosida, {\sf Functional Analysis}, (Springer, Berlin, 1978), Chapt. XI;\\
L. Schwartz, {\sf Th\'eorie des Distributions}, Vol. I,II, (Hermann \& Cie, Paris, 1951);\\
V.S. Vladimirov, {\sf Gleichungen der Mathematischen Physik}, (DVW, Berlin, 1972); (Nauka,
Moscow, 1967).
%--------------------------------------------------------------------------------
%
%[31]
\bibitem{Blumlein:1997em}
  J.~Bl\"umlein and A.~Vogt,
  %``The Evolution of unpolarized singlet structure functions at small x,''
  Phys.\ Rev.\ D {\bf 58} (1998) 014020
%  doi:10.1103/PhysRevD.58.014020
  [hep-ph/9712546].
  %%CITATION = doi:10.1103/PhysRevD.58.014020;%%
%-----------------------------------------------------------------------------------
%
%[32]
\bibitem{Ellis:1993rb}
  R.K.~Ellis, Z.~Kunszt and E.M.~Levin,
  %``The Evolution of parton distributions at small x,''
  Nucl.\ Phys.\ B {\bf 420} (1994) 517--549
   Erratum: [Nucl.\ Phys.\ B {\bf 433} (1995) 498].
%  doi:10.1016/0550-3213(94)90076-0, 10.1016/0550-3213(94)00514-F
  %%CITATION = doi:10.1016/0550-3213(94)90076-0, 10.1016/0550-3213(94)00514-F;%%
%-----------------------------------------------------------------------------------
%
%[33]
\bibitem{Blumlein:2000wh}
  J.~Bl\"umlein, V.~Ravindran and W.L.~van Neerven,
  %``On the Drell-Levy-Yan relation to O(alpha(s)**2),''
  Nucl.\ Phys.\ B {\bf 586} (2000) 349--381
%  doi:10.1016/S0550-3213(00)00422-3
  [hep-ph/0004172].
  %%CITATION = doi:10.1016/S0550-3213(00)00422-3;%%
%------------------------------------------------------------------------
%
%[34]
\bibitem{Duhr:2020seh}
  C.~Duhr, F.~Dulat and B.~Mistlberger,
  {\it The Drell-Yan cross section to third order in the strong coupling constant},
  arXiv:2001.07717 [hep-ph].
  %%CITATION = ARXIV:2001.07717;%%
%------------------------------------------------------------------------
%
%[35]
\bibitem{Blumlein:2012bf}
  J.~Bl\"umlein,
  %``The Theory of Deeply Inelastic Scattering,''
  Prog.\ Part.\ Nucl.\ Phys.\  {\bf 69} (2013) 28--84
%  doi:10.1016/j.ppnp.2012.09.006
  [arXiv:1208.6087 [hep-ph]].
  %%CITATION = doi:10.1016/j.ppnp.2012.09.006;%%
%-------------------------------------------------------------------------------------
%
%[36]
\bibitem{Vermaseren:1998uu}
J.~Vermaseren, { Int. J. Mod. Phys.} {\bf A14} (1999) 2037--2076 {[hep-ph/9806280]}.
%%CITATION = HEP-PH/9806280;%%.
%-------------------------------------------------------------------------------------
%
%[37]
\bibitem{Blumlein:1998if}
J.~Bl{\"u}mlein and S.~Kurth, { Phys. Rev.} {\bf D60} (1999) 014018 {[hep-ph/9810241]}.   
%%CITATION = HEP-PH/9810241;%%.
%-------------------------------------------------------------------------------------
%
%[38]
\bibitem{Ablinger:2013cf}
J.~Ablinger, J.~Bl{\"u}mlein, and C.~Schneider, { J. Math. Phys.} {\bf 54}
  (2013) 082301 {[arXiv: 1302.0378 [math-ph]]}.
%%CITATION = ARXIV:1302.0378;%%.
%-------------------------------------------------------------------------------------
%
%[39]
\bibitem{Ablinger:2014bra}
  J.~Ablinger, J.~Bl\"umlein, C.G.~Raab and C.~Schneider,
  %``Iterated Binomial Sums and their Associated Iterated Integrals,''
  J.\ Math.\ Phys.\  {\bf 55} (2014) 112301
%  doi:10.1063/1.4900836
  [arXiv:1407.1822 [hep-th]].
  %%CITATION = doi:10.1063/1.4900836;%%
%-------------------------------------------------------------------------------------
%
%[40]
\bibitem{Ablinger:2014rba}
J.~Ablinger, PoS (LL2014) 019 {[arXiv:1407.6180[cs.SC]]}.
%%CITATION = ARXIV:1407.6180;%%.
%-------------------------------------------------------------------------------------
%
%[41]
\bibitem{Ablinger:2010kw}
 J.~Ablinger, {\it A Computer Algebra Toolbox for Harmonic Sums Related to
  Particle Physics}, Diploma Thesis, JKU Linz, 2009, arXiv:1011.1176[math-ph].
%-------------------------------------------------------------------------------------
%
%[42]
\bibitem{Ablinger:2013hcp}
 J.~Ablinger, {\it Computer Algebra Algorithms for Special Functions in
  Particle Physics}, Ph.D. Thesis, Linz U. (2012) arXiv:1305.0687[math-ph].
%-------------------------------------------------------------------------------------
%
%[43]
\bibitem{Ablinger:2011te} 
J.~Ablinger, J.~Bl{\"u}mlein, and C.~Schneider, { J. Math. Phys.} {\bf 52}
  (2011) 102301 {[arXiv: 1105.6063
  [math-ph]]}.
%%CITATION = ARXIV:1105.6063;%%.
%-------------------------------------------------------------------------------------
%
%[44]
\bibitem{Ablinger:2017Mellin}
J.~Ablinger, PoS (RADCOR2017) 001 [arXiv:1801.01039 [cs.SC]].
%-------------------------------------------------------------------------------------
%
%[45]
\bibitem{Gross:1973id}
  D.J.~Gross and F.~Wilczek,
  %``Ultraviolet Behavior of Nonabelian Gauge Theories,''
  Phys.\ Rev.\ Lett.\  {\bf 30} (1973) 1343-1346.
%  doi:10.1103/PhysRevLett.30.1343
  %%CITATION = doi:10.1103/PhysRevLett.30.1343;%%
%-----------------------------------------------------------------------------------
%
%[46]
\bibitem{Politzer:1973fx}
  H.D.~Politzer,
  %``Reliable Perturbative Results for Strong Interactions?,''
  Phys.\ Rev.\ Lett.\  {\bf 30} (1973) 1346--1349
%  doi:10.1103/PhysRevLett.30.1346
  %%CITATION = doi:10.1103/PhysRevLett.30.1346;%%
%-----------------------------------------------------------------------------------
%
%[47]
\bibitem{Caswell:1974gg}
  W.E.~Caswell,
  %``Asymptotic Behavior of Nonabelian Gauge Theories to Two Loop Order,''
  Phys.\ Rev.\ Lett.\  {\bf 33} (1974) 244--246.
%  doi:10.1103/PhysRevLett.33.244
  %%CITATION = doi:10.1103/PhysRevLett.33.244;%%
%-----------------------------------------------------------------------------------
%
%[48]
\bibitem{Vladimirov:1979zm}
  A.A.~Vladimirov,
  %``Method for Computing Renormalization Group Functions in Dimensional Renormalization Scheme,''
  Theor.\ Math.\ Phys.\  {\bf 43} (1980) 417--422
   [Teor.\ Mat.\ Fiz.\  {\bf 43} (1980) 210];\\
%  doi:10.1007/BF01018394
  %%CITATION = doi:10.1007/BF01018394;%%
%\bibitem{Baker:1969an}
  M.~Baker and K.~Johnson,
  %``Quantum electrodynamics at small distances,''
  Phys.\ Rev.\  {\bf 183} (1969) 1292--1299.
%  doi:10.1103/PhysRev.183.1292
  %%CITATION = doi:10.1103/PhysRev.183.1292;%%
%-----------------------------------------------------------------------------------
%
%[49]
\bibitem{Blumlein:1994kw}
  J.~Bl\"umlein and J.~Botts,
  %``Do deep inelastic scattering data favor a light gluino?,''
  Phys.\ Lett.\ B {\bf 325} (1994) 190--196
  Erratum: [Phys.\ Lett.\ B {\bf 331} (1994) 450]
%  doi:10.1016/0370-2693(94)91081-2, 10.1016/0370-2693(94)90091-4
  [hep-ph/9401291].
  %%CITATION = doi:10.1016/0370-2693(94)91081-2, 10.1016/0370-2693(94)90091-4;%%
%-----------------------------------------------------------------------------------
%
%[50]
\bibitem{Buza:1996wv}
  M.~Buza, Y.~Matiounine, J.~Smith and W.~L.~van Neerven,
  %``Charm electroproduction viewed in the variable flavor number scheme versus fixed order perturbation theory,''
  Eur.\ Phys.\ J.\ C {\bf 1} (1998) 301--320
%  doi:10.1007/BF01245820
  [hep-ph/9612398].
  %%CITATION = doi:10.1007/BF01245820;%%
%-----------------------------------------------------------------------------------
%
%[51]
\bibitem{BBK2}
%\bibitem{Bierenbaum:2009mv}
  I.~Bierenbaum, J.~Bl{\"u}mlein and S.~Klein,
  %{\sf Mellin Moments of the {$O(\alpha_s^3$)} Heavy Flavor Contributions to
  %unpolarized Deep-Inelastic Scattering at $Q^2 \gg m^2$ and Anomalous
  %Dimensions}, 
  Nucl.\ Phys.\  B {\bf 820} (2009) 417,
   [hep-ph/0904.3563];\\
  %%CITATION = ARXIV:0904.3563;%%
%\bibitem{KleinPHDThesis}
%\cite{Klein:2009ig}
%\bibitem{Klein:2009ig}
  S.~Klein,
  {\it Mellin moments of heavy flavor contributions to $F_2(x,Q^2)$ at NNLO}, PhD Thesis, TU Dortmund, 
  2009, 
  [arXiv:0910.3101 [hep-ph]].
%------------------------------------------------------------------------
%
%[52]
\bibitem{LR}
%\bibitem{Lee:2012cn}
  R.N.~Lee,
  {\it Presenting LiteRed: a tool for the Loop InTEgrals REDuction},
  arXiv:1212.2685 [hep-ph];
  %%CITATION = ARXIV:1212.2685;%%
% \bibitem{Lee:2013mka}
%  R.~N.~Lee,
  %``LiteRed 1.4: a powerful tool for reduction of multiloop integrals,''
  J.\ Phys.\ Conf.\ Ser.\  {\bf 523} (2014) 012059
%  doi:10.1088/1742-6596/523/1/012059
  [arXiv:1310.1145 [hep-ph]].
  %%CITATION = doi:10.1088/1742-6596/523/1/012059;%%
%-----------------------------------------------------------------------------------
%
%[53]
\bibitem{Ablinger:2014vwa}
J.~Ablinger,
A.~Behring, J.~Bl\"umlein, A.~De Freitas, A.~Hasselhuhn, A.~von Manteuffel, M.~Round, C.~Schneider and 
F. Wi\ss{}brock 
  %``The 3-Loop Non-Singlet Heavy Flavor Contributions and Anomalous Dimensions for the Structure Function $F_2(x,Q^2)$ and Transversity,''
  Nucl.\ Phys.\ B {\bf 886} (2014) 733--823
%  doi:10.1016/j.nuclphysb.2014.07.010
  [arXiv:1406.4654 [hep-ph]].
  %%CITATION = doi:10.1016/j.nuclphysb.2014.07.010;%%
%-----------------------------------------------------------------------------------
%
%[54]
\bibitem{Blumlein:2018cms}
  J.~Bl\"umlein and C.~Schneider,
  %``Analytic computing methods for precision calculations in quantum field theory,''
  Int.\ J.\ Mod.\ Phys.\ A {\bf 33} (2018) no.17,  1830015
%  doi:10.1142/S0217751X18300156
  [arXiv:1809.02889 [hep-ph]].
  %%CITATION = doi:10.1142/S0217751X18300156;%%
%-----------------------------------------------------------------------------------
%
%[55]
\bibitem{Ablinger:2018zwz}
  J.~Ablinger, J.~Bl\"umlein, P.~Marquard, N.~Rana and C.~Schneider,
  %``Automated Solution of First Order Factorizable Systems of Differential Equations in One Variable,''
  Nucl.\ Phys.\ B {\bf 939} (2019) 253--291
%  doi:10.1016/j.nuclphysb.2018.12.010
  [arXiv:1810.12261 [hep-ph]].
  %%CITATION = doi:10.1016/j.nuclphysb.2018.12.010;%%
%-----------------------------------------------------------------------------------
%
%[56]
\bibitem{Mertig:1998vk}
  R.~Mertig and R.~Scharf,
  %``TARCER: A Mathematica program for the reduction of two loop propagator integrals,''
  Comput.\ Phys.\ Commun.\  {\bf 111} (1998) 265--273
%  doi:10.1016/S0010-4655(98)00042-3
  [hep-ph/9801383].
  %%CITATION = doi:10.1016/S0010-4655(98)00042-3;%%
%-----------------------------------------------------------------------------------
%
%[57]
\bibitem{Moch:2001zr}
  S.~Moch, P.~Uwer and S.~Weinzierl,
  %``Nested sums, expansion of transcendental functions and multiscale multiloop integrals,''
  J.\ Math.\ Phys.\  {\bf 43} (2002) 3363--3386
%  doi:10.1063/1.1471366
  [hep-ph/0110083].
  %%CITATION = doi:10.1063/1.1471366;%%
%-------------------------------------------------------------------------------------
%
%[58]
\bibitem{Blumlein:2003gb}
  J.~Bl\"umlein,
  %``Algebraic relations between harmonic sums and associated quantities,''
  Comput.\ Phys.\ Commun.\  {\bf 159} (2004) 19--54
  [arXiv:hep-ph/0311046].
  %%CITATION = CPHCB,159,19;%%
%-----------------------------------------------------------------------------------
%
%[59]
\bibitem{SIG1}
C.~Schneider, {S\'em.~Lothar. Combin.\/} {\bf 56} (2007) 1--36
 article B56b.
%-----------------------------------------------------------------------------------
%
%[60]
\bibitem{SIG2}
C.~Schneider, Simplifying Multiple Sums in Difference Fields, in:~{{\sf Computer
Algebra in Quantum Field Theory: Integration,
  Summation and Special Functions}\/} Texts and Monographs in Symbolic
  Computation eds. C.~Schneider and J.~Bl\"umlein  (Springer, Wien, 2013) 325--360
  [arXiv:1304.4134 [cs.SC]].
%-----------------------------------------------------------------------------------
%
%[61]
\bibitem{EMSSP}
%\bibitem{Ablinger:2010pb}
  J.~Ablinger, J.~Bl\"umlein, S.~Klein and C.~Schneider,
  %``Modern Summation Methods and the Computation of 2- and 3-loop Feynman Diagrams,''
  Nucl.\ Phys.\ Proc.\ Suppl.\  {\bf 205-206} (2010) 110--115
  [arXiv:1006.4797 [math-ph]];\\
  %%CITATION = ARXIV:1006.4797;%%
%\bibitem{Blumlein:2012hg}
  J.~Bl\"umlein, A.~Hasselhuhn and C.~Schneider,
  %``Evaluation of Multi-Sums for Large Scale Problems,''
  PoS (RADCOR 2011) 032
  [arXiv:1202.4303 [math-ph]];\\
  %%CITATION = ARXIV:1202.4303;%%
  C. Schneider, % Symbolic Summation in Difference Fields and Its Application in Particle Physics.
  Computer Algebra Rundbrief {\bf 53} (2013) 8--12;\\
  %\bibitem{Schneider:2013zna}
  C.~Schneider,
  %``Modern Summation Methods for Loop Integrals in Quantum Field Theory: The Packages Sigma, EvaluateMultiSums and SumProduction,''
  J.\ Phys.\ Conf.\ Ser.\  {\bf 523} (2014) 012037
  [arXiv:1310.0160 [cs.SC]].
  %%CITATION = ARXIV:1310.0160;%%
%-----------------------------------------------------------------------------------
%
%[62]
\bibitem{PDG}
M. Tanabashi et al. (Particle Data Group), Phys. Rev. D {\bf 98} (2018) 030001 and 2019 update.
%-------------------------------------------------------------------------------------
%
%[63]
\bibitem{SDEP}
%\bibitem{Berends:1987bg}
  F.A.~Berends, G.~Burgers, W.~Hollik and W.L.~van Neerven,
  %``The Standard $Z$ Peak,''
  Phys.\ Lett.\ B {\bf 203} (1988) 177--182;\\
%  doi:10.1016/0370-2693(88)91593-6
  %%CITATION = doi:10.1016/0370-2693(88)91593-6;%%
%\bibitem{Bardin:1988xt}
  D.Y.~Bardin, A.~Leike, T.~Riemann and M.~Sachwitz,
  %``Energy Dependent Width Effects in e+ e- Annihilation Near the Z Boson Pole,''
  Phys.\ Lett.\ B {\bf 206} (1988) 539--542;\\
%  doi:10.1016/0370-2693(88)91627-9
  %%CITATION = doi:10.1016/0370-2693(88)91627-9;%%
%\bibitem{Beenakker:1988pv}
  W.~Beenakker and W.~Hollik,
  %``The Width of the Z Boson,''
  Z.\ Phys.\ C {\bf 40} (1988) 141--148.
%  doi:10.1007/BF01559728
  %%CITATION = doi:10.1007/BF01559728;%%
%-------------------------------------------------------------------------------------
%
%[64]
\bibitem{DENT}
D.~d'Enterria, Slides: Higgs Couplings '17, Heidelberg Nov. 10, 2017;\\ {\tt http://dde.web.cern.ch/}.
%-------------------------------------------------------------------------------------
%
%[65]
\bibitem{Jadach:2019huc}
  S.~Jadach and M.~Skrzypek,
  %``Theory challenges at future lepton colliders,''
  Acta Phys.\ Polon.\ B {\bf 50} (2019) 1705--1717
%  doi:10.5506/APhysPolB.50.1705
  [arXiv:1911.09202 [hep-ph]];\\
S.~Jadach, {\it QED challenges at FCC-ee precision measurements},
PoS (RADCOR2019) 045, file missing; {\tt https://indico.cern.ch/event/783212/timetable/{\#}all.detailed}
%-------------------------------------------------------------------------------------
%
%[66]
\bibitem{Gehrmann:2001pz}
  T.~Gehrmann and E.~Remiddi,
  %``Numerical evaluation of harmonic polylogarithms,''
  Comput.\ Phys.\ Commun.\  {\bf 141} (2001) 296--312
%  doi:10.1016/S0010-4655(01)00411-8
  [hep-ph/0107173].
  %%CITATION = doi:10.1016/S0010-4655(01)00411-8;%%
%-----------------------------------------------------------------------------------
%
%[67]
\bibitem{Ablinger:2018sat}
  J.~Ablinger, J.~Bl\"umlein, M.~Round and C.~Schneider,
  %``Numerical Implementation of Harmonic Polylogarithms to Weight w = 8,''
  Comput.\ Phys.\ Commun.\  {\bf 240} (2019) 189--201
%  doi:10.1016/j.cpc.2019.02.005
  [arXiv:1809.07084 [hep-ph]].
  %%CITATION = doi:10.1016/j.cpc.2019.02.005;%%
%-----------------------------------------------------------------------------------
%
%[68]
\bibitem{FACTS}
N.~Nielsen, {\sf Handbuch der Theorie der Gammafunktion}, (Teubner, Leipzig, 1906); reprinted
by Chelsea Publishing Company, Bronx, NY, 1965;\\
E.~Landau, {\it \"Uber die Grundlagen der Theorie der Fakult\"atenreihen}, S.-Ber. math.-naturw. Kl.
Bayerische Akad. Wiss. M\"unchen, {\bf 36} (1906) 151--218;\\
N.E.~N\"orlund,~{\sf Vorlesungen \"uber Differenzenrechnung}, (Springer, Berlin, 1924); 
reprinted by Chelsea Publishing Company, Bronx, NY, 1954;\\
K.~Knopp, {\sf Theorie und Anwendung der unendlichen Reihen}, (Springer, Berlin, 1947).
%-----------------------------------------------------------------------------------
%
%[69]
\bibitem{ANCONT}
%\bibitem{Blumlein:2000hw}
  J.~Bl\"umlein,
  %``Analytic continuation of Mellin transforms up to two loop order,''
  Comput.\ Phys.\ Commun.\  {\bf 133} (2000) 76--104
%  doi:10.1016/S0010-4655(00)00156-9
  [hep-ph/0003100];\\
  %%CITATION = doi:10.1016/S0010-4655(00)00156-9;%%
%-----------------------------------------------------------------------------------
%\bibitem{Blumlein:2005jg}
  J.~Bl\"umlein and S.O.~Moch,
  %``Analytic continuation of the harmonic sums for the 3-loop anomalous dimensions,''
  Phys.\ Lett.\ B {\bf 614} (2005) 53--61
%  doi:10.1016/j.physletb.2005.03.073
  [hep-ph/0503188].
  %%CITATION = doi:10.1016/j.physletb.2005.03.073;%%
%-----------------------------------------------------------------------------------
%
%[70]
\bibitem{Blumlein:2009ta}
  J.~Bl\"umlein,
  %``Structural Relations of Harmonic Sums and Mellin Transforms up to Weight w = 5,''
  Comput.\ Phys.\ Commun.\  {\bf 180} (2009) 2218--2249
%  doi:10.1016/j.cpc.2009.07.004
  [arXiv:0901.3106 [hep-ph]].
  %%CITATION = doi:10.1016/j.cpc.2009.07.004;%%
%-----------------------------------------------------------------------------------
%
%[71]
\bibitem{Blumlein:2006mh}
  J.~Bl\"umlein, A.~De Freitas, W.L.~van Neerven and S.~Klein,
  %``The Longitudinal Heavy Quark Structure Function F**Q anti-Q(L) in the Region Q**2 >> m**2 at O(alpha**3(s)),''
  Nucl.\ Phys.\ B {\bf 755} (2006) 272--285
%  doi:10.1016/j.nuclphysb.2006.08.014
  [hep-ph/0608024].
  %%CITATION = doi:10.1016/j.nuclphysb.2006.08.014;%%
%-----------------------------------------------------------------------------------
%
%[72]
\bibitem{Vermaseren:1994je}
  J.A.M.~Vermaseren,
  %``Axodraw,''
  Comput.\ Phys.\ Commun.\  {\bf 83} (1994) 45--58.
%  doi:10.1016/0010-4655(94)90034-5
  %%CITATION = doi:10.1016/0010-4655(94)90034-5;%%
%-----------------------------------------------------------------------------------
\end{thebibliography}
